\documentclass[twocolumn]{aastex62}
\pdfoutput=1
\usepackage{gensymb,amsmath}
\usepackage{epstopdf}
\usepackage{url}
\usepackage[]{natbib}

\shorttitle{AGN duty cycle in galaxies since z$\sim$3}
\shortauthors{Delvecchio et al.}
\accepted{to ApJ 19/02/2020}
\begin{document}

\title{THE EVOLVING AGN DUTY CYCLE IN GALAXIES SINCE \lowercase{z}$\sim$3 AS ENCODED \\IN THE X-RAY LUMINOSITY FUNCTION}

\correspondingauthor{Ivan Delvecchio}
\email{ivan.delvecchio@cea.fr}

\author{I. Delvecchio}\altaffiliation{Marie Curie Fellow}
\affil{CEA, IRFU, DAp, AIM, Universit\'e Paris-Saclay, Universit\'e Paris Diderot, Sorbonne Paris Cit\'e, CNRS, F-91191 Gif-sur-Yvette, France}
\affil{INAF - Osservatorio Astronomico di Brera, via Brera 28, I-20121, Milano, Italy \& via Bianchi 46, I-23807, Merate, Italy}
\author{E. Daddi}
\affil{CEA, IRFU, DAp, AIM, Universit\'e Paris-Saclay, Universit\'e Paris Diderot, Sorbonne Paris Cit\'e, CNRS, F-91191 Gif-sur-Yvette, France}
\author{J. Aird}
\affil{Department of Physics \& Astronomy, University of Leicester, University Road, Leicester LE1 7RJ, UK}
\author{J. R. Mullaney}
\affil{Department of Physics and Astronomy, The University of Sheffield, Hounsfield Road, Sheffield S3 7RH, UK}
\author{E. Bernhard}
\affil{Department of Physics and Astronomy, The University of Sheffield, Hounsfield Road, Sheffield S3 7RH, UK}
\author{L. P. Grimmett}
\affil{Department of Physics and Astronomy, The University of Sheffield, Hounsfield Road, Sheffield S3 7RH, UK}
\author{R. Carraro}
\affil{Instituto de F\'\i{}sica y Astronom\'\i{}a, Universidad de Valpara\'\i{}so, Gran Breta\~{n}a 1111, Playa Ancha, Valpara\'\i{}so, Chile}
\author{A. Cimatti}
\affil{University of Bologna, Department of Physics and Astronomy (DIFA), Via Gobetti 93/2, I-40129, Bologna, Italy}
\affil{INAF - Osservatorio Astrofisico di Arcetri, Largo E. Fermi 5, I-50125, Firenze, Italy}
\author{G. Zamorani}
\affil{INAF - Osservatorio di Astrofisica e Scienza dello Spazio, via Gobetti 93/3, I-40129,Bologna, Italy}
\author{N. Caplar}
\affil{Department of Astrophysical Sciences, Princeton University, 4 Ivy Ln., Princeton, NJ 08544, USA}
\author{F. Vito}
\affil{Instituto de Astrofisica and Centro de Astroingenieria, Facultad de Fisica, Pontificia Universidad Catolica de Chile,Casilla 306, Santiago 22, Chile}
\affil{Chinese Academy of Sciences South America Center for Astronomy, National Astronomical Observatories, CAS, Beijing 100012,China}
\author{D. Elbaz}
\affil{CEA, IRFU, DAp, AIM, Universit\'e Paris-Saclay, Universit\'e Paris Diderot, Sorbonne Paris Cit\'e, CNRS, F-91191 Gif-sur-Yvette, France}
\author{G. Rodighiero}
\affil{University of Padova, Department of Physics and Astronomy, Vicolo Osservatorio 3, 35122, Padova, Italy}

\begin{abstract} 

We present a new modeling of the X-ray luminosity function (XLF) of Active Galactic Nuclei (AGN) out to z$\sim$3, dissecting the contribution of main-sequence (MS) and starburst (SB) galaxies. For each galaxy population, we convolved the observed galaxy stellar mass (M$_{\star}$) function with a grid of M$_{\star}$--independent Eddington ratio ($\lambda_{\rm EDD}$) distributions, normalised via empirical black hole accretion rate (BHAR) to star formation rate (SFR) relations. Our simple approach yields an excellent agreement with the observed XLF since z$\sim$3. We find that the redshift evolution of the observed XLF can only be reproduced through an intrinsic flattening of the $\lambda_{\rm EDD}$ distribution, and with a positive shift of the break $\lambda^{*}$, consistent with an anti-hierarchical behavior. The AGN accretion history is predominantly made by massive (10$^{10}<$M$_{\star}<$10$^{11}$~M$_{\odot}$) MS galaxies, while SB-driven BH accretion, possibly associated with galaxy mergers, becomes dominant only in bright quasars, at $\log$(L$_{\rm X}$/erg~s$^{-1}$)$>$44.36~+~1.28$\cdot$(1+z). We infer that the probability of finding highly-accreting ($\lambda_{\rm EDD}>$~10\%) AGN significantly increases with redshift, from 0.4\% (3.0\%) at z=0.5 to 6.5\% (15.3\%) at z=3 for MS (SB) galaxies, implying a longer AGN duty cycle in the early Universe. Our results strongly favor a M$_{\star}$-dependent ratio between BHAR and SFR, as BHAR/SFR $\propto$ M$_{\star}^{0.73[+0.22,-0.29]}$, supporting a non-linear BH buildup relative to the host. Finally, this framework opens potential questions on super-Eddington BH accretion and different $\lambda_{\rm EDD}$ prescriptions for understanding the cosmic BH mass assembly. 

\end{abstract}

\keywords{ galaxies: active--- galaxies: evolution---  galaxies: starburst}

\section{Introduction} \label{intro}
One of the most outstanding achievements of modern astrophysics is the discovery that nearly every galaxy hosts a central supermassive black hole (SMBH), with mass M$_{\rm BH}\sim$10$^{6-10}$~M$_{\odot}$ (e.g.~\citealt{Schmidt1963}; \citealt{Lynden-Bell1969}). SMBHs are believed to grow in mass via accretion of cold gas within the galaxy, occasionally shining as Active Galactic Nuclei (AGN, \citealt{Soltan1982}). Although almost all of today's SMBHs are quiescent, several empirical correlations have been found between M$_{\rm BHs}$ and the properties of local galaxy bulges (e.g. \citealt{Kormendy+2013}), interpreted as the outcome of a long-lasting interplay between SMBH and galaxy growth (e.g. \citealt{Magorrian1998}; \citealt{Ferrarese+2000}; \citealt{Gultekin+2009}). 

To explain this, state-of-the-art numerical simulations advocate a two-fold phase of AGN feedback characterised by high radiative (``quasar mode'') and high kinetic (``jet mode'') luminosities, that combined are able to remove or heat up the gas within the galaxy, via outflows and relativistic jets (\citealt{Sanders+1988}; \citealt{Fabian2012}). Both types of AGN feedback are invoked for gradually hampering the star-forming (SF) content of massive (stellar mass M$_{\star}>$10$^{10}$~M$_{\odot}$) galaxies, thus preventing their runaway mass growth (e.g. \citealt{Hopkins+2008}). While observations and models support this AGN-driven ``quenching'' paradigm to explain the color bimodality and M$_{\star}$ function of local massive systems (e.g. \citealt{Morganti+2003}, \citeyear{Morganti+2005}; \citealt{Fabian2012}; \citealt{Heckman+2014}; \citealt{Benson+2003}; \citealt{Croton+2006}), other studies argue in favor of an AGN-driven enhancement of galaxy star formation rate (SFR, \citealt{Santini+2012}; \citealt{Rosario+2013}; \citealt{Cresci+2015}). 

Though the origin of the SMBH-galaxy co-evolution is not yet fully understood, it is widely accepted that the gas content plays a crucial role in triggering both AGN and star formation activity. Indeed, the SFR is tightly linked to the (molecular) gas content through the Schmidt--Kennicutt relation (hereafter SK relation; \citealt{Schmidt1959}; \citealt{Kennicutt1998}). In parallel, radiative AGN activity (i.e. in the X-rays) is observed to be more prevalent in gas-rich, SF galaxies (e.g. \citealt{Vito+2014}), which might explain the observed positive correlations between SFR and average black hole accretion rate (BHAR, e.g. \citealt{Mullaney+2012}). However, still unclear is whether major mergers or secular processes (e.g. violent disk instabilities, minor mergers) are the leading actors in regulating the growth of SMBHs at different luminosities.

Two main modes of star formation are known to control the growth of galaxies: a relatively steady, secular mode in disk-like galaxies, defining a tight star-forming ``main sequence'' (MS, \citealt{Noeske+2007}; \citealt{Elbaz+2011}; \citealt{Speagle+2014}; \citealt{Schreiber+2015}) between SFR and M$_{\star}$ (1$\sigma$ dispersion of 0.3~dex); and a ``starbursting'' mode above the MS, which is interpreted as driven by mergers \citep{Cibinel+2019}. This latter class of starburst (SB) galaxies is usually defined as showing SFR at least 4$\times$ above the MS, at fixed M$_{\star}$ (e.g. \citealt{Rodighiero+2011}).

Furthermore, multiple studies corroborated the idea that the cold gas fraction f$_{\rm gas}$ (i.e. the ratio between cold gas mass and total baryonic mass, M$_{\rm gas}$/[M$_{\rm gas}$ + M$_{\star}$]) undergoes a strong redshift evolution (f$_{\rm gas}\propto$(1+z)$^2$) in MS galaxies from the local Universe to z$\sim$2 (\citealt{Leroy+2008}; \citealt{Daddi+2010a}; \citealt{Tacconi+2010}; \citealt{Geach+2011}; \citealt{Saintonge+2013}), with a plateau at higher redshift (z$\sim$3, \citealt{Magdis+2013}). At fixed M$_{\rm gas}$, SB galaxies are characterised by higher SFRs compared to MS galaxies, implying higher star formation efficiencies (SFE = SFR/M$_{\rm gas}$, \citealt{Daddi+2010b}; \citealt{Genzel+2010}).

In this context, \citet{Sargent+2012} found that MS and SB galaxies display a bimodal distribution in their specific-SFR (sSFR = SFR/M$_{\star}$), with SB systems contributing to 8--14\% of the total SFR density, up to z$\sim$2. The luminosity threshold above which SB activity dominates the infrared (IR) LF evolves with redshift in a similar fashion with the sSFR of MS galaxies (as $\propto$ (1+z)$^{2.9-3.8}$ with the slope depending on M$_{\star}$), suggesting a roughly constant bimodality at least up to z$\sim$2. 

While both galaxy populations are required to reproduce the total IR (8--1000$~\mu$m) luminosity function, several studies pointed out intrinsic differences between MS and SB galaxies, in terms of structural and physical properties. At z$\sim$0, MS galaxies are preferentially regular disks and less disturbed compared to SB galaxies, which are instead more compact and mostly identified as merging systems, particularly Ultra-Luminous IR Galaxies (ULIRGs, i.e. having IR luminosity L$_{\rm IR}>$10$^{12}$~L$_{\odot}$, e.g. \citealt{Veilleux+2002}). At intermediate redshifts (z$\sim$0.7), \citet{Calabro+2019} observed an increasing incidence of SF clumps when moving above the MS relation, which might indicate a prevalence of merger-induced clumpy star formation toward higher sSFRs. At z$\sim$2, the morphological dichotomy seen in the local Universe becomes much less pronounced, since the fraction of irregular and disturbed morphologies is generally high, and spread out quite uniformly across the SFR--M$_{\star}$ plane (e.g. \citealt{Elmegreen+2007}; \citealt{Forster-Schreiber+2009}; \citealt{Kocevski+2012}). 

Despite much progress having been made in characterising the star formation, gas content, size and morphology between MS and SB galaxies, still unclear is their separate contribution to the global SMBH accretion history. 

Whether AGN activity and star formation evolve in a similar fashion between MS and SB galaxies is still a metter of debate (see \citealt{Rodighiero+2019}). A seminal study of \citet{Mullaney+2012} put forward the idea that the BHAR/SFR ratio is both redshift and M$_{\star}$-invariant, at M$>$10$^{10}$~M$_{\odot}$ and 0.5$<$z$<$2.5. This ``hidden AGN main sequence'' lies at BHAR/SFR$\sim$10$^{-3}$, thus calling for a constant M$_{\rm BH}$/M$_{\star}$ ratio over cosmic time, which would naturally explain the observed M$_{\rm BH}$--M$_{\rm bulge}$ relation at z$\sim$0 \citep{Kormendy+2013}. Lately, other studies have argued in favor of a M$_{\star}$-dependent BHAR/SFR ratio (\citealt{Rodighiero+2015}; \citealt{Yang+2018}; \citealt{Aird+2019}; \citealt{Bernhard+2019}, \citealt{Carraro+2020}), suggesting that BHAR is enhanced relative to SFR in the most massive galaxies. 

Testing whether AGN accretion behaves differently between galaxies on and above the MS relation requires to dissect the X-ray AGN luminosity function (XLF) into those two galaxy classes, and study how they evolve through cosmic time.

This work aims to constrain the relative contribution of MS and SB galaxies to the XLF, since z$\sim$3. In order to avoid selection biases that might arise from collecting AGN at a particular wavelength and/or from flux-limited samples, we model the XLF as the convolution between the galaxy M$_{\star}$ function and a large set of Eddington ratio ($\lambda_{\rm EDD}$) distributions that mimics the stochastic nature of AGN activity (e.g. \citealt{Aird+2013}; \citealt{Conroy+2013}; \citealt{Caplar+2015}; \citealt{Jones+2017}; \citealt{Weigel+2017}; \citealt{Jones+2019}). 
Previous works attempting to achieve this goal \citep{Bernhard+2018} successfully reproduced the observed XLF \citep{Aird+2015} out to z$\sim$1.75, by assuming a M$_{\star}$-dependent shape of the $\lambda_{\rm EDD}$ distribution for SF and quiescent galaxies with relatively complex shapes.

In this work we tackle a simpler approach, showing that a \textit{M$_{\star}$-independent} shape of the $\lambda_{\rm EDD}$ distribution, scaled with a \textit{M$_{\star}$-dependent} normalisation, is fully able to reproduce the observed XLF out to z$\sim$3. This method strongly reduces the number of free parameters, while being fully motivated by recent observational grounds (Section \ref{approach}). Moreover, we are able to predict the relative incidence of AGN of a given L$_{\rm X}$ and redshift, separately within MS and SB galaxies, putting constraints on the typical SMBH duty cycle on and above the MS. This analysis serves as an important test case for making predictions on the expected SMBH growth rate at different redshift, M$_{\star}$ and MS offset.

The manuscript is structured as follows: Section~\ref{method} illustrates our initial assumptions and the statistical approach adopted in this work. The best  XLF prediction for MS and SB galaxies is presented in Section~\ref{results}, quantifying its uncertainties and dissecting its evolution with redshift and L$_{\rm X}$. We further infer the relative contribution of MS and SB galaxies to the global SMBH accretion history since z$\sim$3. In Section~\ref{discussion} we test our modeling, interpret our findings and discuss the implications of this study in the framework of SMBH--galaxy evolution since z$\sim$3. Finally, we list our concluding remarks in Section~\ref{summary}. 

Throughout this paper, we adopt a \citet{Chabrier2003} initial mass function (IMF) and we assume a flat cosmology with $\Omega_{\rm m}$=0.30, $\Omega _{\Lambda}$=0.70 and H$_{0}$=70~km~s$^{-1}$~Mpc$^{-1}$.

\begin{figure*}
     \includegraphics[width=\linewidth]{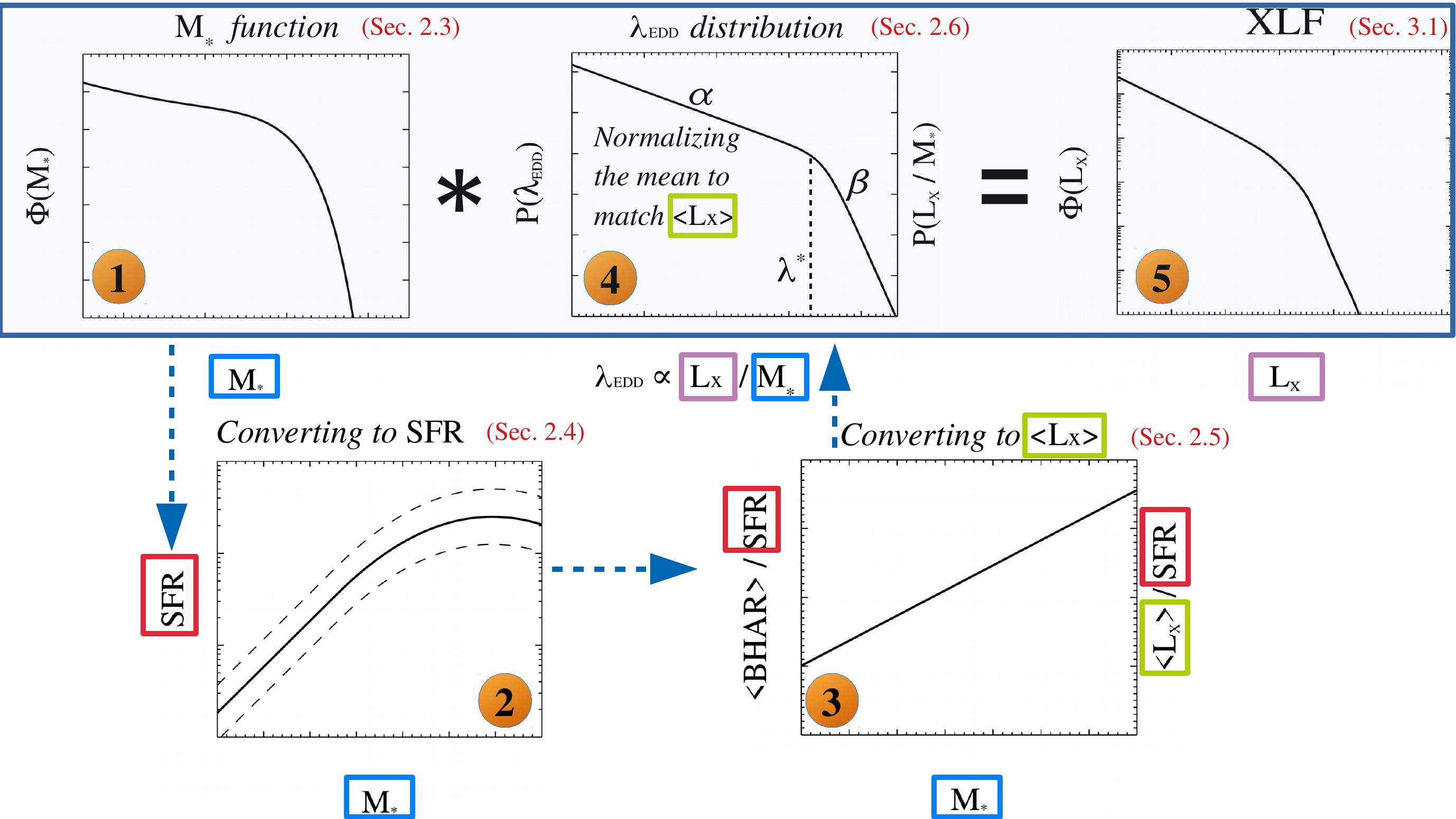}
 \caption{\small Sketch of the convolution model used in this work to derive the XLF. The five steps are summarised as follows. (1): we parametrise the galaxy M$_{\star}$ function of SF galaxies at each redshift (0.5$<$z$<$3). (2): at each M$_{\star}$ and redshift, we read the corresponding SFR from a log-normal SFR kernel centered at the mean MS or SB relation. (3): we derive the expected average $\langle$BHAR$\rangle$ (or $\langle$L$_{\rm X}\rangle$) from M$_{\star}$-dependent BHAR/SFR relations. (4): we assume a large set of $\lambda_{\rm EDD}$ distributions, each normalised to match the corresponding mean $\langle$L$_{\rm X}\rangle$ based on (3) at a given M$_{\star}$ and redshift. (5): each simulated $\lambda_{\rm EDD}$ distribution is convolved with the M$_{\star}$ function (as highlighted in the blue box), yielding the predicted XLF. Each step is described in the corresponding Section, and run separately for MS and SB galaxies. Our predicted XLF (combining MS and SB galaxies) will then be compared with the observed XLF of \citet{Aird+2015} in Section \ref{xlf}. See text for details.
 }
   \label{fig:sketch}
\end{figure*}
  
\section{Methodology} \label{method}

The main goal of the present work is to infer how the average BHAR evolves relative to the host-galaxy mass and star formation activity, while matching the observed evolution of the X-ray emission. This analysis further enables us to constrain the occurrence of AGN activity in galaxies since z$\sim$3, and to dissect the relative contribution of MS and SB populations to the global XLF.

\subsection{Prior assumptions} \label{assumptions} 
Our analysis relies on three prior assumptions, which are listed below.

(i)~ Firstly, we assume that the X-ray AGN LF is predominantly made by MS and SB galaxies. Passive systems, meant to be galaxies well below the MS relation, are assumed to have a negligible contribution ($<$10\%) at all redshifts and at all L$_{\rm X}$. Thus, hereafter we refer to the combined (MS+SB) XLF as ``total XLF''. Here below we report a number of evidence and quantitative arguments supporting our hypothesis. 

The lesser role of quiescent galaxies in the XLF is suggested by studies of the nuclear properties of early-type galaxies, both at z$\sim$0 (e.g. \citealt{Pellegrini+2010}) and at z$\sim$2 (\citealt{Olsen+2013}; \citealt{Civano+2014}). These works generally found low level X-ray AGN activity, with L$_{\rm X}<$10$^{43}$~erg~s$^{-1}$, and predominantly attributed to free-free emission from hot (T$\sim$10$^{6-7}$~K) virialised gas in the galaxy halo \citep{Kim+2013}.
Our assumption is also supported by the significantly smaller reservoirs of cold gas measured in passive galaxies compared to those observed in typical galaxies on the MS relation, despite an important redshift increase at least up to z$\sim$1.5 (see \citealt{Gobat+2018}). Another justification comes from the prevalence of radio AGN within massive and passive galaxies at z$<$1.4 (\citealt{Hickox+2009}; \citealt{Goulding+2014}), which display systematically lower $\lambda_{\rm EDD}$ ($<$10$^{-3}$) than X-ray and MIR-selected AGN ($>$10$^{-2}$). This is also supported by studies on the intrinsic $\lambda_{\rm EDD}$ distribution in quiescent vs. star-forming galaxies, reporting systematically lower mean $\lambda_{\rm EDD}$ values in quiescent systems (e.g. \citealt{Wang+2017}; \citealt{Aird+2019}). Finally, it is worth noticing that the number density of passive galaxies notably drops at z$>$1 (\citealt{Davidzon+2017}), therefore mitigating the incidence of this population at high redshift. Though we acknowledge that passive galaxies might display substantial X-ray emission from hot ionised gas, in this paper we focus on the X-ray emission directly attributed to SMBH accretion. 

A quantitative estimate of the sub-dominant role of quiescent galaxies to the global XLF is presented in Section \ref{qxlf}. Briefly, we conservatively assumed that quiescent galaxies follow the same intrinsic $\lambda_{\rm EDD}$ distribution of MS galaxies at each redshift. Then we re-scaled the $\lambda_{\rm EDD}$ distribution to match empirical mean L$_{\rm X}$/M$_{\star}$ measurements for the quiescent population \citep{Carraro+2020}. This enabled us to quantify upper limits on the space density and luminosity density of quiescent galaxies, confirming their negligible contribution across the L$_{\rm X}$ and redshift range explored in this study.

(ii)~Secondly, we assume that the intrinsic $\lambda_{\rm EDD}$ distribution of AGN follows a broken-powerlaw profile, as parameterised in a number of recent studies (e.g. \citealt{Caplar+2015}, \citeyear{Caplar+2018}; \citealt{Weigel+2017}; \citealt{Bernhard+2018}). This will be further motivated in Section~\ref{eddratio}.
 
(iii)~Lastly, we assume that the faint-end ($\alpha$) and bright-end ($\beta$) slopes of the $\lambda_{\rm EDD}$ distributions do not differ between MS and SB galaxies (Section~\ref{eddratio}), with only the corresponding break values ($\lambda_{\rm EDD}^{*}$) and normalisations being allowed to vary. As a consequence of this assumption, the only free parameter allowed to vary independently among the two populations is $\lambda_{\rm EDD}^{*}$. The main reason is that a simple shift in $\lambda_{\rm EDD}^{*}$ between MS and SB galaxies resembles the well-known double-Gaussian sSFR profile seen in the two populations (e.g. \citealt{Rodighiero+2011}; \citealt{Sargent+2012}). More specifically, since SB-driven star formation is vigorous but short-living (relative to galaxy lifecycle), it does not primarily drive the growth of galaxy M$_{\star}$. Similarly, we can easily parametrise BH accretion in SBs as having much larger BHAR fluctuations compared to the variation of the cumulative BH mass. In addition, our simplistic treatement of SBs is motivated by the un-necessarily high number of free parameters that would otherwise be allowed to vary simultaneously, leading to large degeneracies and poor constraints on the overall behaviour of the $\lambda_{\rm EDD}$ function. We briefly discuss the effect of relaxing this condition in Section~\ref{list}. More details on the $\lambda_{\rm EDD}$ profiles for the MS and SB populations are given in Section~\ref{eddratio}.

\subsection{Our approach} \label{approach}

Our approach is schematically summarised in Fig.~\ref{fig:sketch}, and consists of five steps. We here briefly overview each of them, while a detailed description is presented in the corresponding Sections. 
 
\begin{enumerate}
 \item We parametrise the galaxy M$_{\star}$ function of SF galaxies at each redshift (0.5$<$z$<$3).
 \item At each M$_{\star}$ and redshift, we assign the corresponding SFR by randomly extracting each value from a log-normal SFR kernel, for both MS and SB galaxies, in order to account for the dispersion of the corresponding locus.
\item We derive the expected mean $\langle$BHAR$\rangle$ (or mean $\langle$L$_{\rm X}\rangle$) by multiplying the SFR by various M$_{\star}$-dependent BHAR/SFR relations from the literature.
\item We assume a large set of $\lambda_{\rm EDD}$ distributions, each normalised to have a mean value equal to the corresponding $\langle$L$_{\rm X}\rangle$ expected from (3), at a given M$_{\star}$ and redshift.
\item Each simulated $\lambda_{\rm EDD}$ distribution is convolved with the M$_{\star}$ function, yielding the predicted XLF. These five steps are run separately for MS and SB galaxies. Each predicted XLF (after combining both MS and SB galaxies) is then compared with the observed XLF of \citet{Aird+2015}, as detailed in Section~\ref{results}.
 
\end{enumerate}

In the following Sections, we expand each of the above steps in more detail. A comprehensive list of the free parameters adopted in this work is given in Section~\ref{list} and Table \ref{tab:par}, in which we also discuss the effect induced by each assumption.

\begin{figure}
     \includegraphics[width=\linewidth]{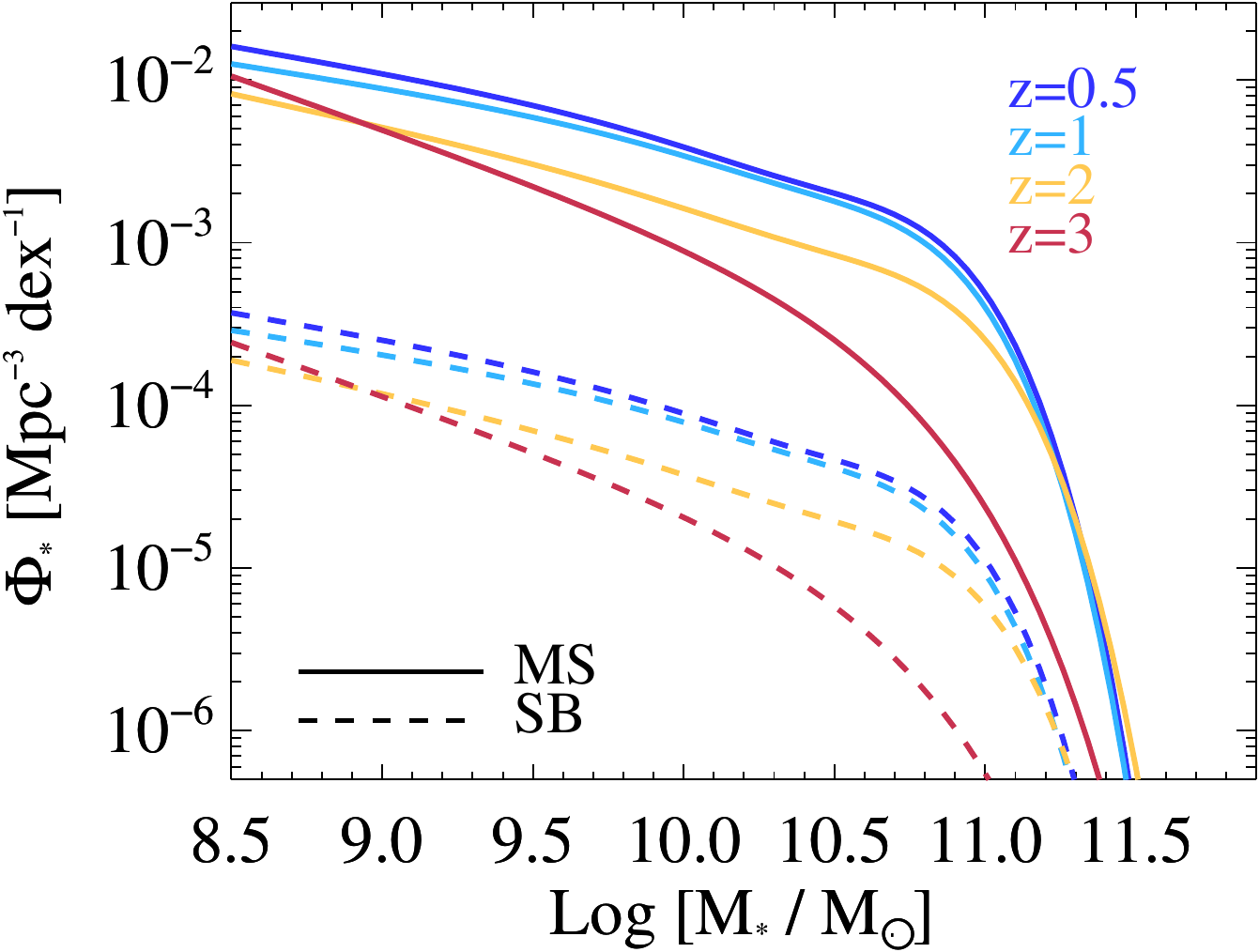}
 \caption{\small Galaxy M$_{\star}$ function at various redshifts for SF galaxies, taken from \citet{Davidzon+2017}. Solid and dashed lines mark MS and SB galaxies, respectively.
 }
   \label{fig:smf}
\end{figure}
  
\subsection{The galaxy stellar mass function}  \label{smf}

The first step displayed in Fig.~\ref{fig:sketch} consists in setting the input galaxy M$_{\star}$ function at different redshifts. The prescription is taken from \citealt{Davidzon+2017}), who exploited the latest optical to infrared photometry collected in the COSMOS field over the UltraVISTA area (1.5~deg$^2$, see \citealt{Laigle+2016}). They provide the M$_{\star}$ function separately between SF and passive galaxies, based on the $[NUV-r]$/$[r-J]$ colors (\citealt{Ilbert+2013}). Throughout this work, we consider only the M$_{\star}$ function relative to the SF galaxy population (i.e. MS and SB galaxies). 

Next, we split the M$_{\star}$ function of SF galaxies among the MS and SB populations. Given that the relative fraction of the two populations has been shown not to vary with M$_{\star}$ (e.g. \citealt{Rodighiero+2011}; \citealt{Sargent+2012}), we only consider how their relative fraction evolves with redshift, by following the prescription of \citet{Bethermin+2012}. The fraction of SB galaxies (f$_{\rm SB}$) appears to evolve linearly, from f$_{\rm SB}$=0.015 (z=0) to f$_{\rm SB}$=0.03 (z=1), while it stays flat at higher redshifts. By simply scaling the SF galaxy M$_{\star}$ function down by f$_{\rm SB}$ (or $1-$f$_{\rm SB}$), we end up with the M$_{\star}$ function of MS and SB galaxies, at each redshift (Fig.~\ref{fig:smf}). We note that the input M$_{\star}$ function of \citet{Davidzon+2017} is already corrected for the Eddington bias, that might have some impact at the high-M$_{\star}$ end. After dissecting among MS and SB galaxies, we interpolate the corresponding M$_{\star}$ function at redshift z=0.5, z=1, z=2 and z=3, across a M$_{\star}$-range of 10$^{8}<$M$_{\star}<$10$^{12}$~M$_{\odot}$.

\begin{figure}
     \includegraphics[width=\linewidth]{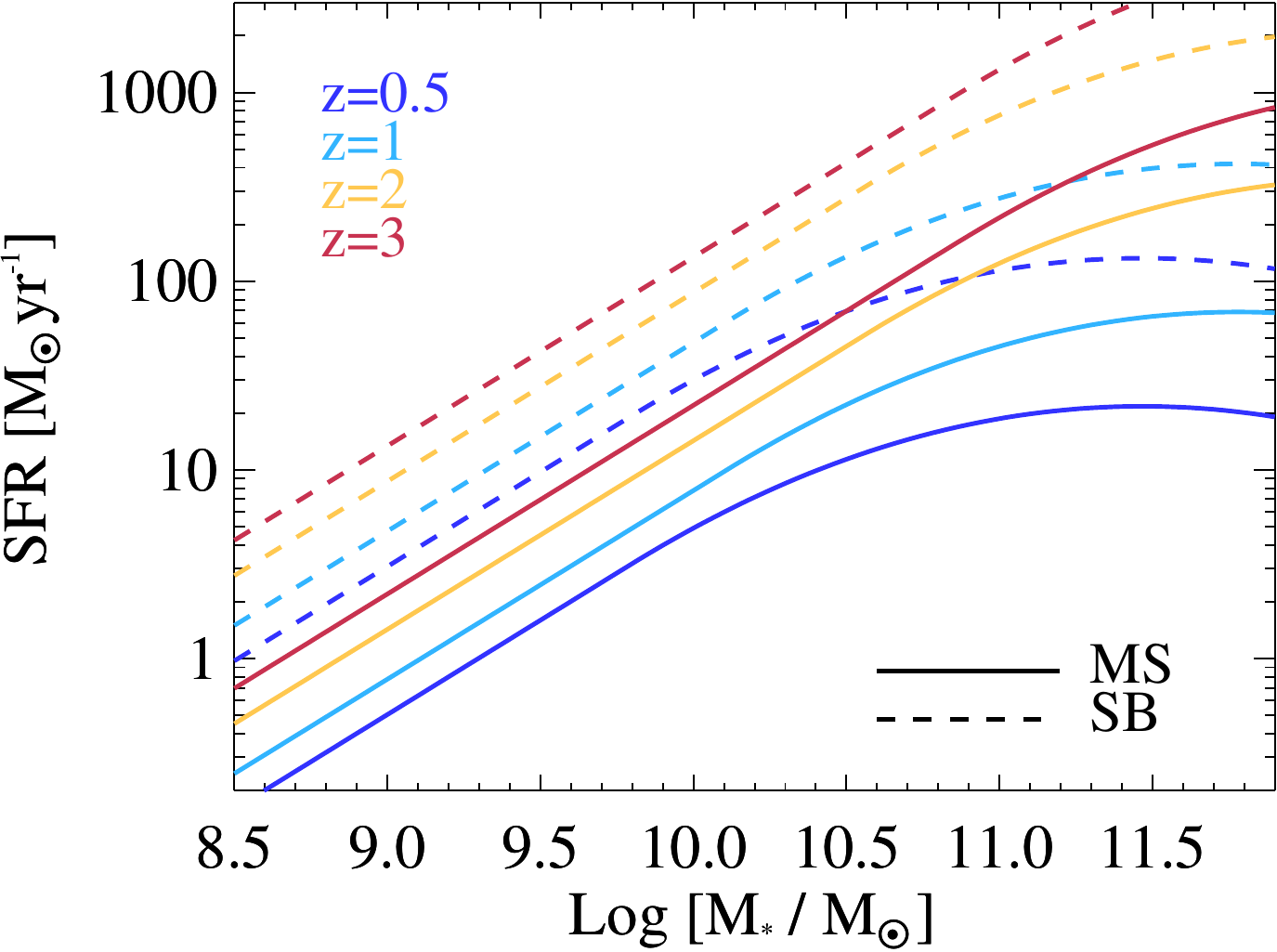}
 \caption{\small Evolving MS relation in the SFR--M$_{\star}$ plane, taken from \citet{Schreiber+2015} and scaled to a \citet{Chabrier2003} IMF. Solid and dashed lines highlight the central locus of MS and SB galaxies, respectively. These were defined as 0.87$\times$SFR$_{\rm MS}$ for MS galaxies, and 5.3$\times$SFR$_{\rm MS}$ for the SB population \citep{Schreiber+2015}. 
 }
   \label{fig:ms_sb}
\end{figure}

\subsection{The MS and the SB loci}  \label{ms_sb}

We use the MS prescription presented in \citet{Schreiber+2015}, which incorporates a redshift evolution and a bending toward higher M$_{\star}$. 

\citet{Schreiber+2015} studied a sample of \textit{Herschel}-selected galaxies out to z$\sim$4, dissecting the observed distribution of MS offset (=SFR/SFR$_{\rm MS}$, see their Eq.~10 and Fig.~19) among the MS and SB populations. By fitting that distribution via a double log-normal function, MS galaxies are centered at 0.87$\times$SFR$_{\rm MS}$, while SB galaxies are centered at 5.3$\times$SFR$_{\rm MS}$ \citep{Schreiber+2015}. Both relations were re-scaled to a \citet{Chabrier2003} IMF. The 1$\sigma$ dispersion for both the MS and SB loci is assumed to be 0.3~dex (e.g. \citealt{Speagle+2014}). We note that such MS relation displays a bending toward the highest M$_{\star}$, which makes the transition from MS to SB galaxies not a linear function of sSFR. The adopted MS is qualitatively similar to other recent M$_{\star}$--dependent prescriptions (e.g. \citealt{Lee+2015}; \citealt{Scoville+2017}), while a single powerlaw MS would deliver slightly higher SFR estimates (e.g. \citealt{Rodighiero+2015}), yet consistent results within the uncertainties (e.g. \citealt{Yang+2018}).

At fixed M$_{\star}$ and redshift, we account for the MS dispersion by randomly extracting each SFR from a log-normal SFR kernel centered as described above. This is shown in Fig.~\ref{fig:ms_sb} at various redshifts, and separately for MS (solid lines) and SB (dashed lines) galaxies.

\subsection{The BHAR/SFR relation with M$_{\star}$} \label{bhsf}

As shown in Fig.~\ref{fig:sketch}, the third step consists of converting the derived SFR into BHAR. A number of BHAR--SFR relations have been proposed in the literature, mostly relying on X-ray and IR observations of AGN samples (e.g. \citealt{Shao+2010}; \citealt{Rosario+2012}; \citealt{Page+2012}; \citealt{Mullaney+2012}; \citealt{Chen+2013}; \citealt{Delvecchio+2015}; \citealt{Stanley+2015}).

In order to mitigate possible selection biases induced by short-term ($<$1~Myr, \citealt{Schawinski+2015}) AGN variability, an effective approach is starting from large M$_{\star}$-selected samples, and averaging AGN activity over galaxy timescales ($>$100~Myr) to unveil the ``typical'' SMBH accretion rate across the full galaxy lifecycle (\citealt{Hickox+2014}). 

This approach was first pioneered in \citet{Mullaney+2012}, who used a $Ks$-selected galaxy sample to investigate the BHAR/SFR relationship in the GOODS-South field. Interestingly, they found a roughly constant BHAR/SFR$\sim$10$^{-3}$ with redshift (at 0.5$<$z$<$2.5), which nicely reproduced the local M$_{\rm BH}$--M$_{\rm bulge}$ correlation as the consequence of steady SMBH accretion and SF activity over cosmic time \citep{Kormendy+2013}. 

\begin{figure}
     \includegraphics[width=\linewidth]{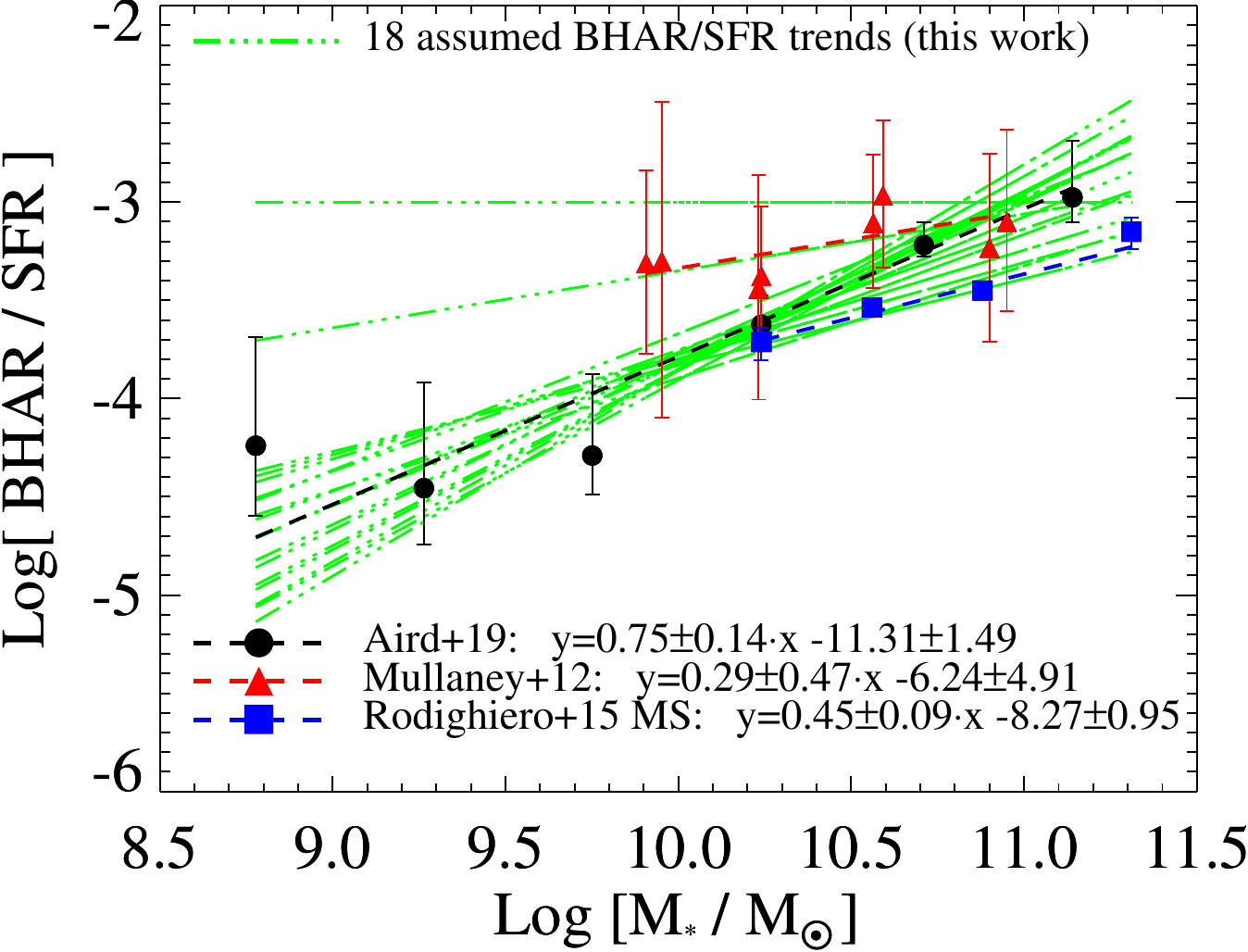}
 \caption{\small Relationship between BHAR and SFR, as a function of M$_{\star}$. Dashed lines represent a least squares fitting (in the $\log$-$\log$ space) of the data presented in \citeauthor{Mullaney+2012} (\citeyear{Mullaney+2012}, red triangles), \citeauthor{Rodighiero+2015} (\citeyear{Rodighiero+2015}, blue squares) and \citeauthor{Aird+2019} (\citeyear{Aird+2019}, black circles), respectively. The best-fit slope and normalisation of each fit are reported in the legend. The green dot-dashed lines indicate the 18 different BHAR/SFR relations explored in this work: 15 of them are taken within 2$\sigma$ around the prescription of \citet{Aird+2019}, while the remaining three are taken to match the \citeauthor{Rodighiero+2015}, \citeauthor{Mullaney+2012}, and the flat BHAR/SFR=10$^{-3}$ trends. See Section~\ref{bhsf} for more details.
 }
   \label{fig:bhar_sfr}
\end{figure}

Moreover, \citet{Rodighiero+2015} analysed $BzK$-selected galaxies at z$\sim$2, split between MS, SB and passive systems \citep{Daddi+2004}, in the COSMOS field \citep{Scoville+2007}. The authors found that MS galaxies display a M$_{\star}$-dependent BHAR/SFR relation, as BHAR/SFR $\propto$ M$_{\star}^{0.44}$. In addition, they argued that SB galaxies at z$\sim$2 show 2$\times$ lower BHAR/SFR ratios relative to MS analogs at the same M$_{\star}$. 

More recently, \citet{Aird+2019} adopted a Bayesian approach to reconstruct the intrinsic $\lambda_{\rm EDD}$ distribution across the full galaxy population: they corroborated the need for a linearly M$_{\star}$-dependent BHAR/SFR at 0.5$<$z$<$2.5, roughly independent on redshift and on galaxy sSFR. \citet{Delvecchio+2015} explored the average BHAR/SFR in a sample of \textit{Herschel}-selected galaxies at z$<$0.5, finding no obvious difference as moving above the MS relation. From those studies, it is therefore still unclear whether such BHAR/SFR relation evolves with M$_{\star}$ and redshift, and whether SB galaxies are truly offset from this trend.

Given these open questions, we prefer to explore a wide set of BHAR/SFR correlations, at each redshift, spanning the full parameter space of slopes and normalisations probed by previous studies. Specifically, 15 different slopes have been explored around the most recent derivation of \citet{Aird+2019}, plus three additional ones to account for different results (Fig.~\ref{fig:bhar_sfr}, green dot-dashed lines). These 18 slopes are chosen as follows: 15 are uniformly extracted within 2$\sigma$ from the most recent derivation by \citet{Aird+2019}; the remaining 3 are taken to match the relationships found by \citet{Mullaney+2012}, \citet{Rodighiero+2015}, and a flat BHAR/SFR=10$^{-3}$ as the most extreme case. For each slope around the best-fit of \citet{Aird+2019}, the relative normalisation is set accordingly to fit the corresponding data-points of Fig.~\ref{fig:bhar_sfr}. Therefore, slope and normalisation of each relation are co-variant and count as a single free parameter.

We explore the full set of BHAR/SFR relations \textit{at each redshift}, assuming that MS and SB galaxies share the same trend, since no stringent constraints on a potential deviation are clearly found in the literature (e.g. \citealt{Delvecchio+2015}; \citealt{Yang+2018}; \citealt{Aird+2019}). Although we acknowledge that some studies argued in favor of a 2$\times$ lower BHAR/SFR in SB at z=2, relative to MS galaxies, we caution that a substantial BHAR contribution might be highly obscured, especially in compact SB galaxies at high redshift, and unaccounted-for via a simple hardness ratio technique (\citealt{Aird+2015}; \citealt{Bongiorno+2016}). We note that a positive redshift dependence was claimed in \citet{Yang+2018}, who assumed a single powerlaw MS relation (from \citealt{Behroozi+2013}) at each redshift. However, if taking a bending MS toward high M$_{\star}$, especially at lower redshifts (e.g. \citealt{Schreiber+2015}, \citealt{Scoville+2017}) we remark that all previous studies are consistent with a redshift-invariant BHAR/SFR ratio.

For each (M$_{\star}$,z,SFR), the resulting BHAR is simply calculated by multiplying the BHAR/SFR ratio by the corresponding SFR obtained from Section~\ref{ms_sb}. We stress that such BHAR is meant to be the ``mean'' linear BHAR ($\langle$BHAR$\rangle$ hereafter). This is connected to the mean X-ray luminosity $\langle L_{\rm X} \rangle$ as follows:
\begin{equation}
 \langle L_{\rm X} \rangle = \frac{\epsilon ~ c^2}{1 - \epsilon} \cdot \frac{\langle \rm BHAR \rangle}{k_{\rm BOL}}
 \label{eq:lx}
\end{equation}
where \textit{c} is the speed of light in the vacuum, \textit{$\epsilon$} is the matter-to-radiation conversion efficiency, and k$_{\rm BOL}$ is the [2--10]~keV bolometric correction. If assuming $\epsilon$=0.1 (e.g. \citealt{Marconi+2004}; \citealt{Hopkins+2007}) and a single k$_{\rm BOL}$=22.4 (median value found by \citealt{Vasudevan+2007} in local AGN samples), Eq.~\ref{eq:lx} reduces to $\langle L_{\rm X}$/[erg~s$^{-1}$]$\rangle$ = 2.8$\times$10$^{44} \langle$BHAR/$[M_{\odot}$yr$^{-1}]\rangle$.

We acknowledge that the k$_{\rm BOL}$ is known to exhibit a positive $L_{\rm X}$-dependence (e.g. \citealt{Marconi+2004}; \citealt{Lusso+2012}). However, in this study we are not \textit{assuming} a k$_{\rm BOL}$. More simply, in order to scale the average BHAR back to L$_{\rm X}$, we need to adopt the same k$_{\rm BOL}$ value (e.g. 22.4) used in previous works (\citealt{Mullaney+2012}; \citealt{Rodighiero+2015}; \citealt{Aird+2019}), otherwise we would obtain inconsistent L$_{\rm X}$ from what they started. In other words, we used the same k$_{\rm BOL}$ as previous studies to get rid of the k$_{\rm BOL}$ dependence when computing the XLF.

\subsection{The Eddington ratio distribution of AGN} \label{eddratio}

In this work, we express the Eddington ratio $\lambda_{\rm EDD}$ as a proxy for L$_{\rm X}$/M$_{\star}$ (or BHAR/M$_{\star}$), traditionally named ``specific L$_{\rm X}$'' (or ``specific BHAR'', sBHAR). This formalism has been used by many authors to quantify how fast the SMBH is accreting relative to the M$_{\star}$ of the host galaxy (e.g. \citealt{Aird+2012}, \citeyear{Aird+2018}). This quantity is likely more physically meaningful than the absolute L$_{\rm X}$, since it accounts for the bias that a more massive galaxy with a given BH $\lambda_{\rm EDD}$ would appear more luminous than a less massive galaxy at the same $\lambda_{\rm EDD}$.
In particular, assuming a fixed k$_{\rm BOL}$ (=22.4, see Section~\ref{bhsf}) and a fixed M$_{\rm BH}$/M$_{\star}$ ratio of 1/500 \citep{Haring+2004}, the $\lambda_{\rm EDD}$ can be linked to L$_{\rm X}$/M$_{\star}$ via:

\begin{equation}
 \lambda_{\rm EDD} = \rm \frac{k_{\rm BOL} ~ \rm L_{\rm X}}{1.3 \times 10^{38} \times \rm 0.002~M_{\star}/M_{\odot}}
   \label{eq:eddratio}
\end{equation}

We will briefly discuss in Section \ref{ledd_true} how a M$_{\star}$--dependent M$_{\rm BH}$/M$_{\star}$ ratio (see e.g. \citealt{Delvecchio+2019}) would affect the derived $\lambda_{\rm EDD}$ distributions and our global picture of cosmic BH growth.

A single, universal powerlaw shape was firstly proposed by \citet{Aird+2012} by analysing the incidence of X-ray AGN activity in galaxies at 0.2$<$z$<$1.0. The assumption of a broken powerlaw, with a break close to the Eddington limit, has been found to better reproduce the observed shape of the XLF \citep{Aird+2013}. Despite the M$_{\star}$-invariant distribution, they observed a steep redshift increase of its normalisation, as $\propto$(1+z)$^{3.8}$ (see also \citealt{Bongiorno+2012}). 

In order to reproduce the XLF at z$\gtrsim$1, a M$_{\star}$-dependent $\lambda_{\rm EDD}$ distribution has been implemented in a number of studies (e.g. \citealt{Bongiorno+2016}; \citealt{Georgakakis+2017}; \citealt{Jones+2017}; \citealt{Aird+2018}; \citealt{Bernhard+2018}). Building on those findings, we attempt to keep the $\lambda_{\rm EDD}$ distribution as simple as possible, while being consistent with the BHAR/SFR trends reported in the recent literature.

\begin{figure}
     \includegraphics[width=\linewidth]{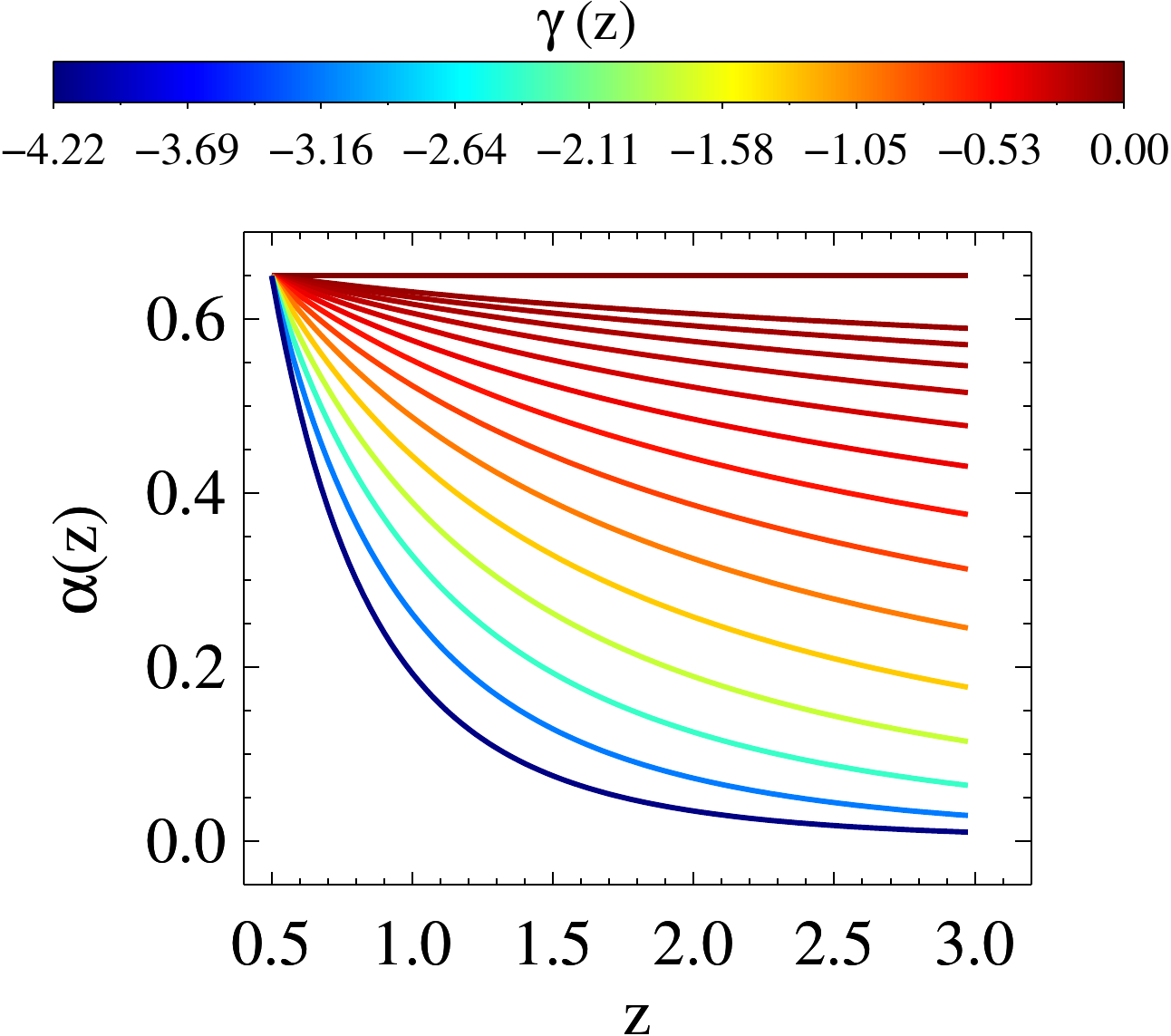}
 \caption{\small colored lines show the 15 $\alpha$ vs redshift trends assumed in this work. Each line marks a different $\gamma$ value, spanning from steep to flat trends with redshift, reaching to $\alpha \sim$0 at z=3 in the most extreme case. We set $\alpha$=0.55 at z=0.5 to be consistent with previous tudies ($\alpha$=0.65$\pm$0.05 at 0.2$<$z$<$1.0, \citealt{Aird+2012}; $\alpha$=0.45, \citealt{Caplar+2018}).
 }
   \label{fig:gamma}
\end{figure}

\subsubsection{Broken powerlaw and flattening with redshift} \label{eddratio_1}

The fourth step of Fig.~\ref{fig:sketch} shows the shape of the assumed $\lambda_{\rm EDD}$ distribution. As mentioned in Section~\ref{method}, we assume a broken powerlaw with faint- and bright-end slopes $\alpha$ and $\beta$, respectively, that meet at the break $\lambda^{*}$. In order to mitigate the parameter degeneracy, we assume that MS and SB galaxies share the same slopes ($\alpha$, $\beta$), while the corresponding breaks $\lambda^{*}_{\rm MS}$ and $\lambda^{*}_{\rm SB}$ are allowed to vary independently within the range [-1; +0.5] in $\log$ space (with steps of 0.1), at each redshift. This $\lambda^{*}$ range was chosen to be consistent with the typical position of the knee found in recent determinations of the $\lambda_{\rm EDD}$ distribution at z$\lesssim$2 (e.g. \citealt{Caplar+2018}; \citealt{Bernhard+2018}; \citealt{Aird+2019}). 

As mentioned in Sect.~\ref{assumptions}, while a double-Gaussian profile describes the sSFR variation between MS and SB galaxies (e.g. \citealt{Rodighiero+2011}; \citealt{Sargent+2012}), similarly a shift in $\lambda_{\rm EDD}^{*}$ enables us to describe the difference in sBHAR (or Eddington ratio) among those populations. In particular, SB galaxies show intense and short-lasting SFR variations relative to MS analogs, thus not important to explain the growth of galaxy M$_{\star}$ (e.g. \citealt{Rodighiero+2011}; \citealt{Sargent+2012}). With such a formalism, we can parametrise BH accretion in SBs as an intense and short-lasting phenomenon too, characterised by much larger BHAR fluctuations compared to the variation of the cumulative BH mass.

We further reduce the parameter space by imposing that $\lambda^{*}_{\rm SB} > \lambda^{*}_{\rm MS}$, in order to reproduce the systematically higher BHAR found in SB galaxies contrained by previous studies (\citealt{Rodighiero+2015}; \citealt{Delvecchio+2015}; \citealt{Aird+2019}; \citealt{Grimmett+2019}). Moreover, both slopes and $\lambda^{*}$ values are assumed to be M$_{\star}$--invariant. Although we acknowledge that the intrinsic $\lambda_{\rm EDD}$ shape might be more complex (M$_{\star}$--dependent, see e.g. \citealt{Bernhard+2018}; \citealt{Aird+2019}; \citealt{Grimmett+2019}) this simplistic prescription allows us to link the evolution of the mean expected $\lambda_{\rm EDD}$ to a rigid shift in $\lambda^{*}$. The only foreseen M$_{\star}$ dependence comes from the adopted BHAR/SFR relation (Section \ref{bhsf}), as described in Section \ref{eddratio_2}.

In addition, we implement a flattening of the faint-end slope $\alpha$ with redshift. This is supported by recent studies attempting at reproducing the AGN bolometric LF (\citealt{Caplar+2015}; \citealt{Weigel+2017}; \citealt{Jones+2019}), by convolving the galaxy M$_{\star}$ function with some M$_{\star}$--dependent p($\log \lambda_{\rm EDD}$) distribution of AGN. Similarly for our XLF, if the faint-end of the M$_{\star}$ function steepens with redshift (Fig.~\ref{fig:smf}), while the faint-end XLF flattens with redshift, this latter feature can be reproduced if the p($\log \lambda_{\rm EDD}$) at low $\lambda_{\rm EDD}$ intrinsically flattens with redshift (\citealt{Bongiorno+2016}; \citealt{Bernhard+2018}; \citealt{Caplar+2018}).

We parametrise the redshift flattening of $\alpha$ as follows:
\begin{equation}
   \alpha(z) ~ =~  \alpha(z_0) \cdot \bigg(\frac{1+z}{1+z_0}\bigg)^{\gamma}
   \label{eq:gamma}
  \end{equation}

For a given $\lambda_{\rm EDD}$ distribution, the bright-end slope $\beta$ is directly linked to the bright-end slope of the AGN bolometric LF \citep{Caplar+2015}, since it is much flatter than the exponential decline of the galaxy M$_{\star}$ function at the high-M$_{\star}$ end. Previous studies found that $\beta$ ranges between 1.8 and 2.5 (\citealt{Hopkins+2007}; \citealt{Caplar+2015}, \citeyear{Caplar+2018}). However, steeper slopes might be still accommodated in case multiple contributions are overimposed one another (i.e. MS and SB). In order to account for this and for a possible redshift evolution of $\beta$, we assume $\beta$ to take the values [2,3,4,5]. As pointed out in \citet{Caplar+2018}, we stress that changing $\beta$ within those values has a negligible impact on the integrated X-ray luminosity density. We therefore anticipate that our analysis is not able to tightly constrain this parameter (see Section~\ref{par_evo}). We refer the reader to Table \ref{tab:par} for an exhaustive list of the aforementioned assumptions.

\begin{figure}
     \includegraphics[width=\linewidth]{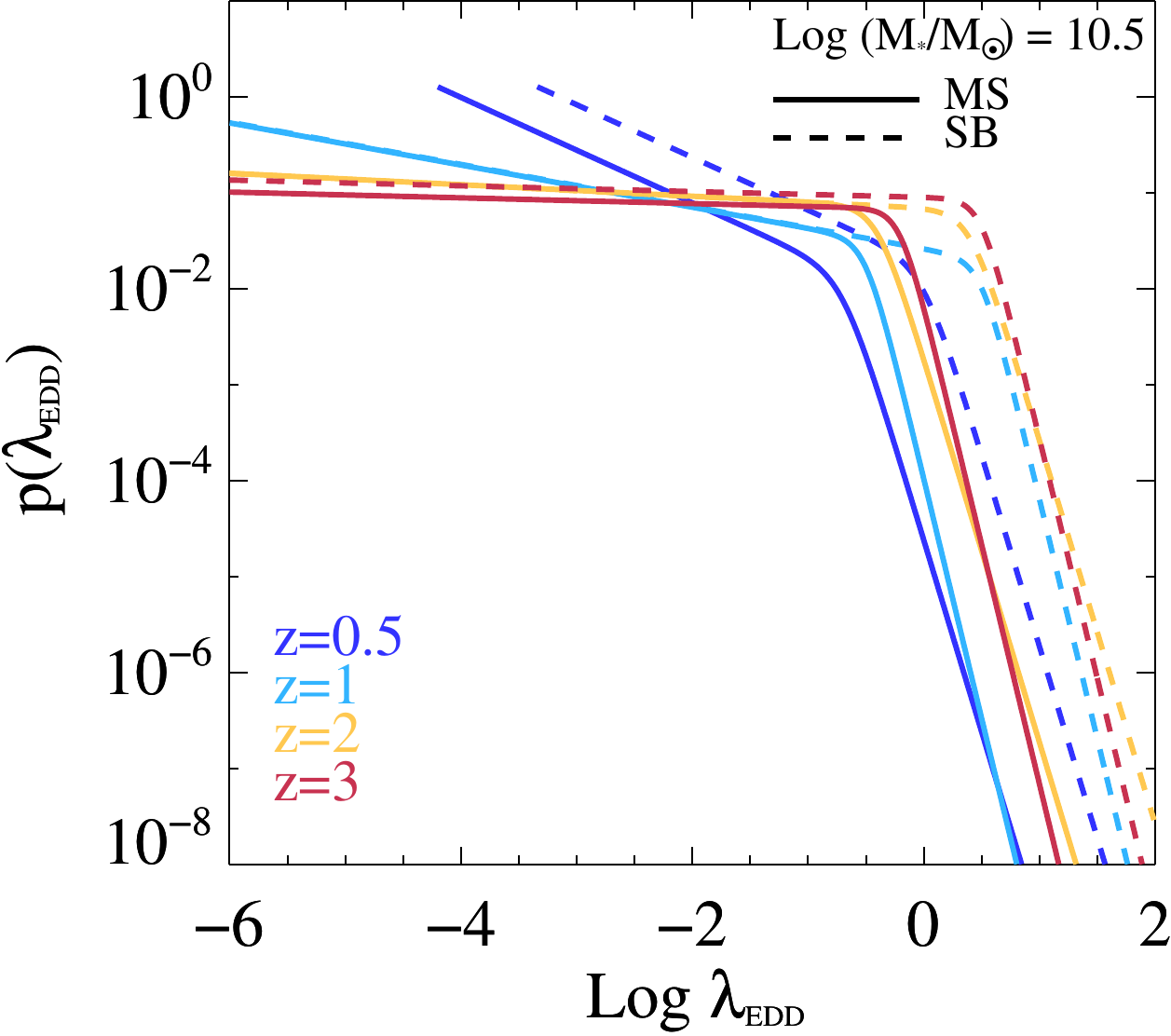}
 \caption{\small Set of $\lambda_{\rm EDD}$ probability distributions that best reproduce the observed XLF (see Section \ref{xlf}), here shown only at M$_{\star}$=10$^{10.5}$~M$_{\odot}$ for illustrative purpose. We indicate MS and SB galaxies with solid and dashed lines, respectively. Colours mark our four redshift bins. All functions are calculated down to their minimum $\lambda_{\rm MIN}$ and normalised to unity, as detailed in Section \ref{eddratio_2}.
 }
   \label{fig:ledd_plot}
\end{figure}

\subsubsection{The probability density function} \label{eddratio_2}

To trace the distribution of $\lambda_{\rm EDD}$, we measure the probability density function p($\log \lambda_{\rm EDD} | $M$_{\star}$,z) as a function of (M$_{\star}$,z), that is defined as follows (see \citealt{Aird+2019}):
\begin{equation}
 \int_{-\infty}^{\infty} p(\log \lambda_{\rm EDD} | M_{\star},z) ~ d \log \lambda_{\rm EDD} ~~ = ~~ 1
\label{eq:norm}
\end{equation}

This approach assumes that \textit{all} SMBHs are accreting, however weak their accretion rate might be. Therefore, p($\log \lambda_{\rm EDD}$) reflects the entire distribution of specific L$_{\rm X}$/M$_{\star}$ encompassed by SMBHs during their life cycle. According to this formalism, the mean $\lambda_{\rm EDD}$ of the model ($\langle \lambda_{\rm mod} \rangle$) defines the ``typical'' $\langle$L$_{\rm X}$/$M_{\star}\rangle$ averaged over the entire SMBH life cycle. This quantity can be written as:
\begin{equation}
  \langle \lambda_{\rm mod} \rangle = \int_{\log \lambda_{\rm MIN}}^{\log \lambda_{\rm MAX}}  \lambda_{\rm EDD} \cdot p(\log \lambda_{\rm EDD}) ~ d \log \lambda_{\rm EDD}
  \label{eq:mean}
\end{equation}

For simplicity, we do not assume a M$_{\star}$-dependent \textit{shape} of the $\lambda_{\rm EDD}$ distribution. Instead, at each (M$_{\star}$,z) we tailor the minimum $\lambda_{\rm EDD}$ ($\lambda_{\rm MIN}$) in order to normalise our p($\log \lambda_{\rm EDD}$) to 1 (Eq.~\ref{eq:norm}), while anchoring the mean $\langle \lambda_{\rm mod} \rangle$ (Eq.~\ref{eq:mean}) to match empirical BHAR/SFR trends, as explained below.

Firstly, at fixed (M$_{\star}$,z), we can set the expected average SFR (Section~\ref{ms_sb}) and the average BHAR (Section~\ref{bhsf}), which yields a mean expected Eddington ratio (or $\langle L_{\rm X} \rangle$), namely $\langle \lambda_{\rm exp} \rangle$. 

Secondly, in order to match $\langle \lambda_{\rm mod} \rangle$ and $\langle \lambda_{\rm exp} \rangle$, we scan each simulated $\lambda_{\rm EDD}$ distribution backwards from the maximum ($\log \lambda_{\rm MAX}$=2), and assuming a logarithmic step $\Delta (\log \lambda_{\rm EDD})$=0.02. At each iteration, we calculate the corresponding $\langle \lambda_{\rm mod} \rangle$ (Eq.~\ref{eq:mean}) and compare it with the expected $\langle \lambda_{\rm exp} \rangle$ taken from Section~\ref{bhsf}. We stop when $\langle \lambda_{\rm mod} \rangle$ equals $\langle \lambda_{\rm exp} \rangle$ within 0.02~dex, which sets $\lambda_{\rm MIN}$. Below this value, we impose p($\log \lambda_{\rm EDD}$)=0. In case $\log \lambda_{\rm MIN}<$-6 (i.e. L$_{\rm X}\lesssim$10$^{40}$~erg~s$^{-1}$), we truncate the $\lambda_{\rm EDD}$ distribution at that value, since current observational data do not probe down to the corresponding L$_{\rm X}$ (Fig. \ref{fig:xlf_aird}). Our arbitrary choice of $\log \lambda_{\rm MAX}$=2 does not impact our procedure, since the distribution drops steeply above the Eddington limit (Section~\ref{list}).

We iterate the procedure described above at each M$_{\star}$, redshift, BHAR/SFR trend, and for every combination of the p($\log \lambda_{\rm EDD}$) parameters: $\alpha$ (or equivalently $\gamma$), $\beta$, $\lambda^{*}_{\rm MS}$ and $\lambda^{*}_{\rm SB}$.

Fig.~\ref{fig:ledd_plot} shows the set of p($\log \lambda_{\rm EDD}$) (for galaxies at M$_{\star}$=10$^{10.5}$~M$_{\odot}$) that best reproduce the observed XLF at different redshifts (see Section \ref{xlf}). Each function is defined down its corresponding $\lambda_{\rm MIN}$ and normalised to unity.

\begin{deluxetable*}{cccc}
\tablecaption{List of free parameters, ranges and relative assumptions made in this work (see also Section~\ref{list}). The motivation behind each assumption is described in the corresponding Sections. We briefly summarise (column 4) the effect produced by each assumption, to help the reader distinguish the genuine trends from those induced by our prior hypotheses. The reference BHAR/SFR trends are taken from \citeauthor{Mullaney+2012} (\citeyear{Mullaney+2012}, M12); \citeauthor{Rodighiero+2015} (\citeyear{Rodighiero+2015}, R15), \citeauthor{Aird+2019} (\citeyear{Aird+2019}, A19).
}
\tablewidth{0pt}
\tablehead{\colhead{Parameter} & \colhead{Range} & \colhead{Assumptions} & \colhead{Effects} }
\decimalcolnumbers
\startdata
$\alpha$ & [0.01; 0.55] & \textbullet{} $\alpha$[z=0.5] = 0.55  &   \textbullet{} reduce the parameter space   \\
    &    & and evolves as (1+z)$^\gamma$  &  in line with empirical studies  \\
    &    & with $\gamma$=[-4.22; 0] &  (Section \ref{list})  \\
   \smallskip  &    & (Section \ref{eddratio_1} and Fig.\ref{fig:gamma}) &      \\
    &    & \textbullet{} $\alpha$(MS) = $\alpha$(SB)  &   \textbullet{} reduce the number of free parameters \\
   \smallskip    &    & (Section \ref{assumptions} and Fig.\ref{fig:ledd_plot}) & that could not be constrained  \\
   \smallskip    &    &     &   \textbullet{} link sSFR and sBHAR variations (Sect.\ref{eddratio_1})    \\
    &    &  \textbullet{} independent of M$_{\star}$   &   \textbullet{} simplify the shape of $p(\log \lambda_{\rm EDD})$  \\
    &    &   (Section \ref{eddratio_1})   &     \\
\hline    
 $\beta$  &  [2,3,4,5]  &  \textbullet{}  $\beta$(MS) = $\beta$(SB) &    \textbullet{} reduce the number of free parameters   \\
      \smallskip  &    & (Section \ref{assumptions} and Fig.\ref{fig:ledd_plot}) &  that could not be constrained    \\ 
         \smallskip    &    &     &   \textbullet{} link sSFR and sBHAR variations (Sect.\ref{eddratio_1})      \\
     &    &  \textbullet{} independent of M$_{\star}$   &   \textbullet{} simplify the shape of $p(\log \lambda_{\rm EDD})$   \\
    &    &   (Section \ref{eddratio_1})   &     \\
\hline
 $\log \lambda^{*}_{\rm MS}$  &  [-1; +0.5[  &  \textbullet{} full range explored at each z     &   \textbullet{} The positive shift with redshift   \\
                    \smallskip       &               &             (Section \ref{eddratio_1})              &  is genuine (Section \ref{par_evo}) \\
                &              &  \textbullet{} independent of M$_{\star}$   &   \textbullet{}  The z-evolution of $\lambda^{*}$ is mirrored in L$_{\rm X}^{*}$ \\
                &              &     (Section \ref{eddratio_1})      &    at each M$_{\star}$ (Section \ref{xlf_mass}) \\
\hline
$\log \lambda^{*}_{\rm SB}$    &  ]-1; +0.5]  &  \textbullet{} $\lambda^{*}_{\rm SB}>\lambda^{*}_{\rm MS}$   &   \textbullet{} The positive shift with redshift   \\
                          \smallskip           &               &                  (Section \ref{eddratio_1})            &  is induced by $\lambda^{*}_{\rm MS}$ (Section \ref{par_evo}) \\
                &              &  \textbullet{} independent of M$_{\star}$     &   \textbullet{} The SB-MS offset in $\lambda^{*}$ is mirrored in L$_{\rm X}^{*}$ \\
                &              &     (Section \ref{eddratio_1})      &    at each M$_{\star}$ (Section \ref{xlf_mass}) \\
\hline
BHAR/SFR                         &  [0; 1.05]  &     \textbullet{} 18 values: 15 around A19   &    \textbullet{} M$_{\star}$--dependent BHAR/SFR ratios    \\
  slope$^{(**)}$ with $\log$ M$_{\star}$ &               &     + 3 to match M12, R15      &  are favoured, but a flat trend is rejected  \\
                                              &               &     and a flat trend at 10$^{-3}$  &  at $\sim$3$\sigma$ (Section \ref{fgas})  \\
                           \smallskip         &               &   (Section \ref{bhsf} and Fig.~\ref{fig:bhar_sfr})   &     \\ 
                                              &               &     \textbullet{} The mean $\langle BHAR \rangle$ anchors  &   \textbullet{} The minimum $\lambda_{\rm MIN}$ changes with M$_{\star}$    \\
                                             &               &      the mean $p(\log \lambda_{\rm EDD})$ value at each M$_{\star}$  & to accommodate a M$_{\star}$--independent  \\
                           \smallskip         &               &    (Section \ref{eddratio_2})     &  $p(\log \lambda_{\rm EDD})$ shape (Section \ref{eddratio_2})   \\
                                              &               &     \textbullet{} same relation for MS and SB  &   \textbullet{}  The constant $\frac{SFR_{\rm SB}}{SFR_{\rm MS}}$ induces a constant  \\
                                              &               &  (Section \ref{bhsf})             &  mean $\frac{\langle BHAR_{\rm SB} \rangle}{\langle BHAR_{\rm MS} \rangle} \approx$0.8~dex, at each M$_{\star}$ and z    \\
                           \smallskip         &               &                         &  (Section \ref{bhard} and Fig.\ref{fig:bhard}) \\
                                 &               &     \textbullet{} full range explored at each z  &  \textbullet{} The non-evolution of this relation\\
                                 &               &                   (Section \ref{bhsf})           &   with redshift is genuine (Section \ref{par_evo})\\
\enddata
\tablecomments{$(**)$ The relative normalisation is chosen to fit the corresponding data-points of Fig \ref{fig:bhar_sfr}. }
\label{tab:par}
\end{deluxetable*}

\subsection{Free parameters} \label{list}

In this Section we summarise the five free parameters introduced in our analysis: ($\alpha$, $\beta$, $\lambda^{*}_{\rm MS}$, $\lambda^{*}_{\rm SB}$ and the BHAR/SFR relation). A comprehensive list of all prior assumptions made for these parameters is detailed in Table \ref{tab:par}. Next to each assumption, we report the effect produced in this work, in order to help the reader distinguish between genuine trends and behaviors obtained by construction.

The faint and bright-end slopes ($\alpha$, $\beta$) of the $\lambda_{\rm EDD}$ distribution are assumed for simplicity to be the same between MS and SB galaxies. Relaxing this condition would increase the parameter degeneracy, without adding constraints on the intrinsic $\alpha$(SB) and $\beta$(MS) at low and high $L_{\rm X}$, respectively.
Specifically, our prior assumption on $\alpha$ consists in a progressive flattening of p($\lambda_{\rm EDD}$) with redshift, in order to reproduce the faint-end flattening of the XLF toward higher redshifts (Section~\ref{eddratio_1}). We start from $\alpha$=0.55 at z=0.5, which is consistent with the faint-end $\lambda_{\rm EDD}$ slope presented in previous studies at z$<$1: ($\alpha$=0.65$\pm$0.05, \citealt{Aird+2012}; $\alpha$=0.45, \citealt{Caplar+2018}). 

The bright-end slope $\beta$ is instead assumed to take the values [2, 3, 4, 5], in order to cover the typical range of bright-end slopes found in the AGN bolometric LF (\citealt{Hopkins+2007}; \citealt{Caplar+2015}, \citeyear{Caplar+2018}).

The break Eddington ratio of MS and SB galaxies ($\lambda^{*}_{\rm MS}$, $\lambda^{*}_{\rm SB}$) are instead let free to vary over the range [-1; +0.5] with a uniform logarithmic step of 0.1. In order to be consistent with recent papers finding systematically higher mean BHAR in SB relative to MS galaxies (\citealt{Rodighiero+2015}; \citealt{Yang+2018}; \citealt{Aird+2019}; \citealt{Grimmett+2019}; \citealt{Carraro+2020}), we accordingly impose that $\lambda^{*}_{\rm SB} > \lambda^{*}_{\rm MS}$.

Finally, the BHAR/SFR slope ranges between 0 and 1.05, covering various empirical trends with M$_{\star}$ reported in the recent literature (Section \ref{bhsf} and Fig.~\ref{fig:bhar_sfr}). Each normalisation is set to minimise the corresponding $\chi^2$.

\section{Results} \label{results}

In this Section we present the results of our modeling to reproduce the XLF since z$\sim$3. The galaxy M$_{\star}$ function (Section~\ref{smf}) was convolved with a set of M$_{\star}$--independent Eddington ratio parameters (slopes and break, see Sections \ref{eddratio_1} and \ref{eddratio_2}), but with a M$_{\star}$--dependent normalisation that matches the mean BHAR from several BHAR/SFR relations found in the literature (Section~\ref{bhsf}). This analysis was run separately among MS and SB galaxies, which allowed us to infer the relative contribution of each class to the total XLF.   
With this formalism, the XLF $\Phi(L_{\rm X},z)$ was derived as follows:

\begin{equation}
 \Phi(L_{\rm X},z) = \int_{M_{\star}}^{} \Phi_{\star}(M'_{\star},z) \circledast p(L_{\rm X} | M'_{\star},z) ~ dM'_{\star} 
\label{eq:phi_lx}
\end{equation}
where $\Phi_{\star}(M'_{\star},z)$ is the galaxy M$_{\star}$ function of the corresponding population, and $p(L_{\rm X} | M'_{\star},z)$ is the likelihood distribution of $\log$~L$_{\rm X}$ as a function of ($M_{\star}$,z). The total XLF split between MS and SB galaxies is shown in Section~\ref{xlf}. The best parameters, along with their uncertainties and confidence ranges are listed in Table \ref{tab:uncertainties}. The degeneracy and the evolution of each free parameter are discussed in Section~\ref{par_evo}, where we also present the SMBH accretion rate density dissected for the first time among those two populations.

\subsection{The total XLF of MS and SB galaxies since z$\sim$3} \label{xlf}
By combining the free parameters listed in Table \ref{tab:par}, we generate 129,600 predicted XLFs in total. This comes by multiplying the following numbers: 15 ($\alpha$ values), 4 ($\beta$ values), 18 (BHAR/SFR trends with $M_{\star}$) and (16$\times$15)/2 combinations of $\lambda^{*}$ (accounting for the condition $\lambda^{*}_{\rm SB} > \lambda^{*}_{\rm MS}$).

Following Eq.~\ref{eq:phi_lx} we calculate the total XLF and compare each derivation with the latest observed XLF presented in \citet{Aird+2015}. The authors separately calculated the XLF both in the soft (0.5--2 keV) and in the hard (2--10 keV) X-ray bands, and combined them consistently in a single dataset at 2--10~keV. They also subtracted the X-ray emission expected from star formation \citep{Aird+2017}, and further corrected for incompletness and AGN obscuration. Therefore this compilation is the most complete XLF constrained by X-ray observations over such a luminosity and redshift range.

\begin{figure*}
     \includegraphics[width=\linewidth]{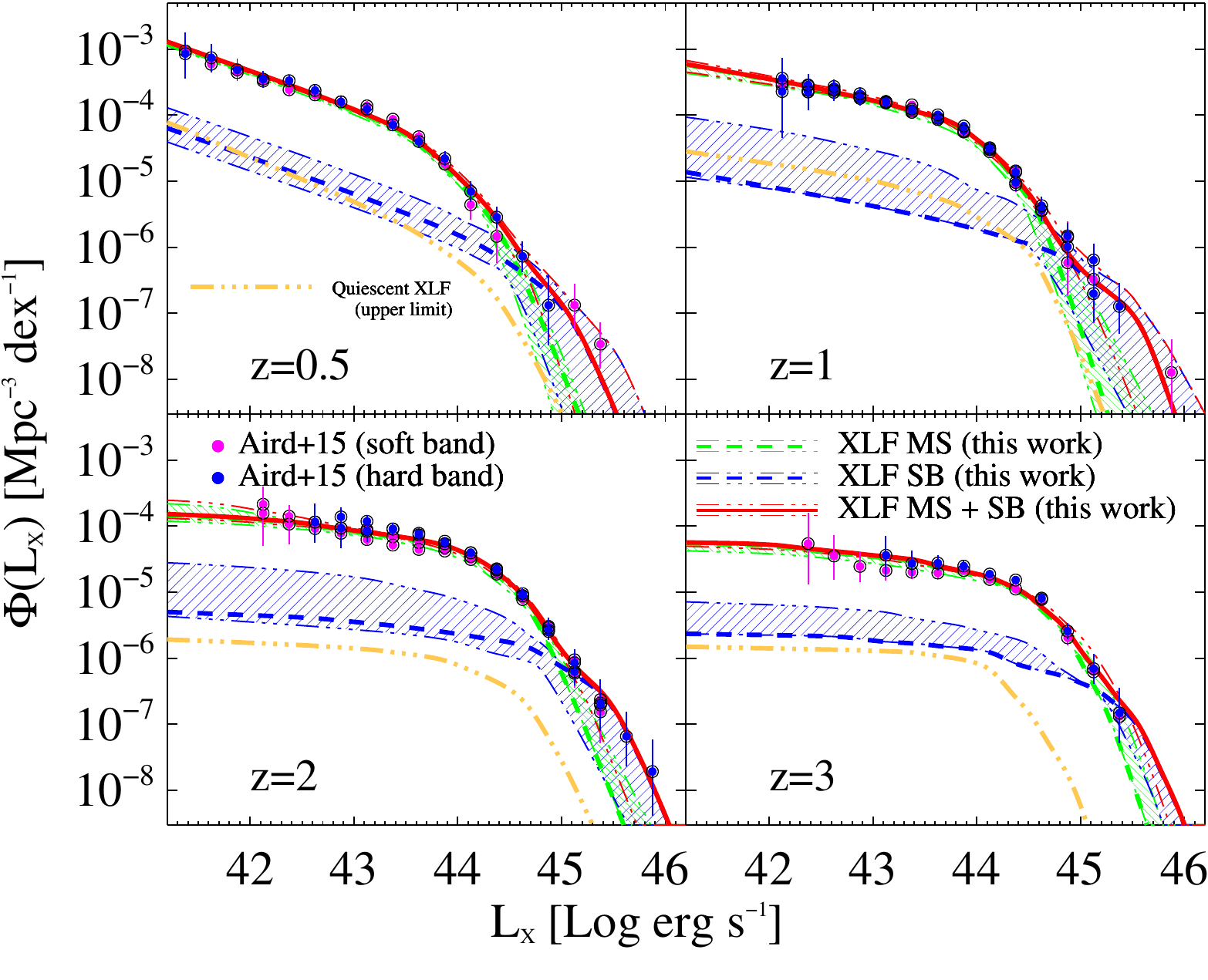}
 \caption{\small The best 2--10 keV AGN X-ray luminosity function (XLF) predicted by our modeling at various redshifts (red solid lines). The XLF is dissected between MS (green dashed lines) and SB (blue dashed lines) galaxies at each redshift, with their $\pm$1$\sigma$ confidence interval being enclosed by the corresponding dot-dashed lines. The upper limit XLF made by quiescent galaxies (orange dotted-dashed line) is detailed in Section \ref{qxlf}. Datapoints are from the compilation of \citet{Aird+2015}, containing both data in the soft (0.5--2 keV, magenta points) and in the hard (2--10 keV, blue points) X-ray bands.   
 }
   \label{fig:xlf_aird}
\end{figure*}

The observed datapoints of \citet{Aird+2015} are given both in the soft (magenta) and in the hard (blue) X-ray bands. Some redshift bins in Fig.~\ref{fig:xlf_aird} display two datasets from \citeauthor{Aird+2015} (e.g. at both 0.8$<$z$<$1.0 and 1.0$<$z$<$1.2 in our z=1 bin), that were taken in order to match the mean redshift between data and our model predictions. 
Among the \citet{Aird+2015} published datapoints, we exploited only those lying at high enough L$_{\rm X}$ where the contamination from galaxy star formation is negligible (see Fig.~8 in \citealt{Aird+2015}), namely $>$10$^{41.3}$~erg~s$^{-1}$ at z=0.5 and $>$10$^{42}$~erg~s$^{-1}$ in the other bins. 

We select the best XLF model prediction via a simple $\chi^2$ minimization. Starting from $\alpha(z=0.5)$ = 0.55, we firstly identify the $\gamma$ value that best describes the observed XLF across all redshift bins (i.e. minimising the global $\chi^2$ at 0.5$\leq$z$\leq$3). This led us to $\gamma$=-3.16$^{+0.79}_{-0.00}$\footnote{Zero errors are due to our discrete grid and should be interpreted as smaller than the closest value (see Table \ref{tab:uncertainties}).} (at 1$\sigma$ level), that defines the flattening trend with redshift. Secondly, among the pool of models within 1$\sigma$ from the best $\gamma$, we searched for the best XLF at each redshift, based on $\chi^2$ minimization. 

Although the comparison with \citet{Aird+2015} does not allow us to test the separate contribution of MS and SB galaxies to the observed XLF, it is important to verify that our combined (MS+SB) XLF agrees with current data. This is not obvious, as we stress again that our best XLF is \textit{not} actually a fit, but the model prediction that best agrees with the observed XLF of \citet{Aird+2015}. Fig.~\ref{fig:xlf_aird} shows the best XLF (red solid lines) at each redshift, and split between the MS (green dashed lines) and SB (blue dashed lines) populations. 

The range of XLFs corresponding to $\pm$1$\sigma$ confidence interval is enclosed within the corresponding dot-dashed lines. Such range is delimited by all the predicted XLFs within a certain $\Delta \chi^2$ threshold with $N_{\rm dof}$ degrees of freedom from the best XLF\footnote{$N_{\rm dof}$ is the difference between the observed datapoints N$_{\rm d}$ and the number of free parameters of each redshift bin (i.e. N$_{\rm d}$--4 at z=0.5, N$_{\rm d}$--5 in the other bins, corresponding to $\Delta \chi^2$=4.71 and 5.88, respectively}). The confidence interval around the best XLF also incorporates the propagation of the uncertainties on $\gamma$.

The agreement with the XLF of \citet{Aird+2015} is striking in all redshift bins, suggesting that our simple statistical approach, constrained by empirical grounds, is able to successfully reproduce the XLF since z$\sim$3 without invoking complex $\lambda_{\rm EDD}$ shapes or large numbers of free parameters. 

As reported in \citet{Aird+2015}, the observed XLF is best reproduced with a flexible double powerlaw (FDPL) model, incorporating both a L$_{\rm X}$--dependent flattening at the faint-end and a positive L$_{\rm X}$ shift with redshift. In our modeling, we also assumed that MS and SB galaxies follow the same intrinsic shape in $\lambda_{\rm EDD}$, independently of M$_{\star}$ (Section \ref{assumptions} and \ref{eddratio_1}). The only difference in $\lambda_{\rm EDD}$ is driven by the corresponding break $\lambda^{*}$. With this simple formalism, the flattening of the XLF with redshift is reproduced through a significant flattening of the $\alpha$ slope (Fig. \ref{fig:gamma}); whereas the L$_{\rm X}$ shift is obtained through a gradual predominance of SB galaxies toward higher L$_{\rm X}$. This feature comes naturally from our initial assumptions that $\lambda_{\rm EDD, SB}>\lambda_{\rm EDD, MS}$, in accordance with the higher specific BHARs found in SB relative to MS galaxies (\citealt{Rodighiero+2015}; \citealt{Delvecchio+2015}; \citealt{Bernhard+2016}; \citealt{Yang+2018}; \citealt{Aird+2019}). This trend also agrees with previous studies (e.g. \citealt{Caplar+2015}, \citeyear{Caplar+2018}) finding that $\lambda^{*}\propto$ 0.048(1+z)$^{2.5}$ at z$<$2, and constant at z$\gtrsim$2 (see Fig.~\ref{fig:par_evo}).

Fig.~\ref{fig:xlf_aird} clearly shows that MS galaxies dominate $\Phi(L_{\rm X})$ at L$_{\rm X} \lesssim $10$^{44.5}$~erg~s$^{-1}$, while SB galaxies tend to take over at higher luminosities, with a cross-over L$_{\rm X}^{\rm cross}$ that slightly evolves with redshift (see Section~\ref{fgas} and Fig.~\ref{fig:par_evo2}).

\begin{figure*}
     \includegraphics[width=\linewidth]{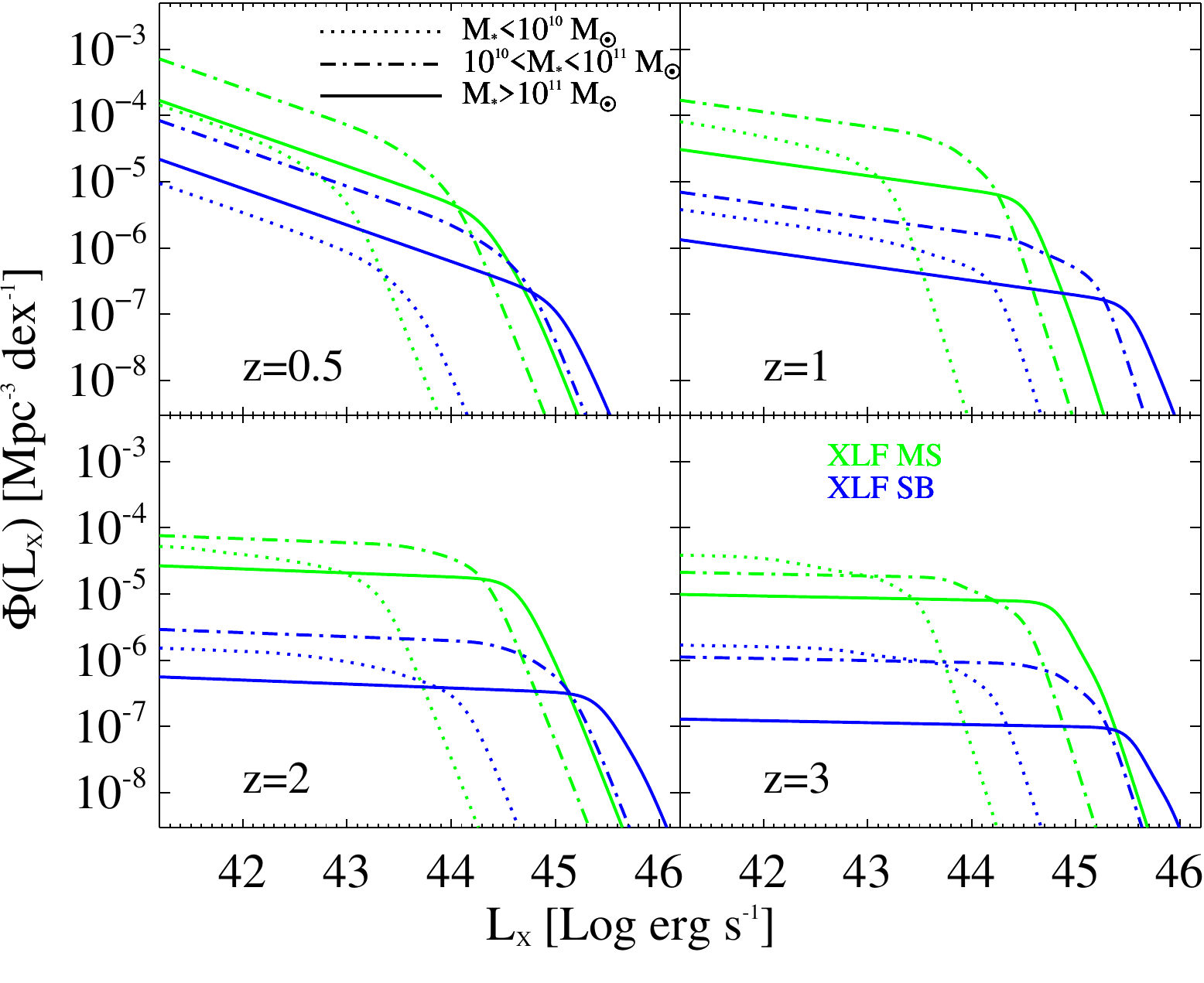}
 \caption{\small The best 2--10 keV AGN X-ray luminosity function (XLF) dissected among three M$_{\star}$ bins: M$_{\star}<$10$^{10}$ (dashed lines), 10$^{10}<$M$_{\star}<$10$^{11}$ (dot-dashed lines) and M$_{\star}>$10$^{11}$~M$_{\odot}$ (solid lines), separately for MS (green) and SB (blue) galaxies.
 }
   \label{fig:xlf_aird_mass}
\end{figure*}

\subsubsection{The sub-dominant contribution of quiescent galaxies to the global XLF} \label{qxlf}

As mentioned in Section \ref{assumptions}, here we quantitatively address the contribution of quiescent galaxies to the global XLF at various redshifts. 
For consistency, we adopt a similar approach to that presented for star-forming galaxies (Section \ref{approach}). We convolve the galaxy M$_{\star}$ function of quiescent galaxies \citep{Davidzon+2017} with a set of $\lambda_{\rm EDD}$ distributions. Instead of plugging in our modeling five additional free parameters for the quiescent galaxy population, we conservatively assume that they share the same best $\lambda_{\rm EDD}$ parameters of MS galaxies ($\alpha$, $\beta$, $\lambda^{*}_{\rm MS}$) inferred from Section \ref{xlf}. The normalisation of the corresponding $\lambda_{\rm EDD}$ distribution is, however, set differently to match empirical studies of quiescent galaxies. We base our validation on recent observational grounds by \citep{Carraro+2020}, who stacked deep \textit{Chandra} images of the COSMOS field, including a M$_{\star}$ complete sample of quiescent galaxies. They inferred mean L$_{\rm X}$ as a function of M$_{\star}$ and redshift out to z$\sim$3, which we use to anchor the mean $\lambda_{\rm EDD}$ of the corresponding distributions (as presented in Section \ref{eddratio_2}). Our convolution yields the XLF of quiescent galaxies out to z$\sim$3 (orange dashed-dotted lines in Fig. \ref{fig:xlf_aird}). While we acknowledge that our approach is not formally the same as that adopted for MS ad SB galaxies, we stress that our reasoning is largely supported by observational studies finding lower mean BHAR (or $\lambda_{\rm EDD}$) in quiescent galaxies compared to star-forming analogs (e.g. \citealt{Rodighiero+2015}; \citealt{Bernhard+2018}; \citealt{Yang+2018}; \citealt{Aird+2019}). We therefore interpret the resulting quiescent XLF as an upper limit. Nevertheless, if the intrinsic $\lambda_{\rm EDD}$ distribution differs from that of MS galaxies, we stress that the normalisation and break L$_{\rm X}$ of the quiescent XLF would change toward opposite directions. Given the generally negligible contribution of quiescent galaxies displayed at all L$_{\rm X}$ (Fig. \ref{fig:xlf_aird}), we might expect them to become potentially comparable to SB galaxies only at low L$_{\rm X}$, where SBs are already sub-dominant and poorly constrained. 
We stress that the mean stacked L$_{\rm X}$ reported by \citeauthor{Carraro+2020} display a positive dependence on both M$_{\star}$ and redshift, in accordance with previous studies (see e.g. \citealt{Wang+2017}; \citealt{Bernhard+2018}; \citealt{Aird+2018}). The sub-dominant role highlighted by this test might be caused by the steep drop of the galaxy quiescent M$_{\star}$ function toward higher redshift, which counter-balances the increasing X-ray AGN fraction (but see \citealt{Georgakakis+2014}). 

It is worth noticing that neglecting the quiescent galaxy population does not contradict the observed prevalence of X-ray AGN with SFR 2$\times$ below the MS (e.g. \citealt{Mullaney+2015}). Indeed, the definition of MS adopted in this work accounts for a 1$\sigma$ scatter of a factor of two (Sect. \ref{ms_sb}). This implies that X-ray AGN lying 2$\times$ below the MS are within the MS locus, and therefore would not nominally contribute to quiescent galaxies. In addition, X-ray AGN found below the MS display Seyfert-like luminosities (L$_{\rm X}<$10$^{44}$~erg~s$^{-1}$), thus consistent with MS galaxies hosting moderately luminous AGN. In summary, our check demonstrates that quiescent galaxies make a very minor ($<$10\%) contribution to the space density of X-ray AGN at all L$_{\rm X}$ and redshifts analysed in this work. We believe this justifies not plugging them in our modeling.

\subsection{XLF split in M$_{\star}$ bins} \label{xlf_mass}

We further explore the differential contribution of galaxies of different M$_{\star}$ to the observed XLF. To do this we dissect our best model prediction in three M$_{\star}$ bins (M$_{\star}<$10$^{10}$, 10$^{10}<$M$_{\star}<$10$^{11}$ and M$_{\star}>$10$^{11}$~M$_{\odot}$), and separately between MS and SB galaxies, as shown in Fig.~\ref{fig:xlf_aird_mass}. We remind the reader that the XLF comes from the convolution of the galaxy M$_{\star}$ function and the $\lambda_{\rm EDD}$ distribution: because we assumed a M$_{\star}$--independent shape of the $\lambda_{\rm EDD}$ and a M$_{\star}$--dependent normalisation, the differential contribution in M$_{\star}$ is mostly driven by the M$_{\star}$-dependent BHAR/SFR ratio. This dictates a shift of the mean $\lambda_{\rm EDD}$ with M$_{\star}$, which translates to a different break L$_{\rm X}$ with M$_{\star}$. 

Given these considerations, unsurprisingly we see that galaxies of different M$_{\star}$ dominate at different L$_{\rm X}$. Particularly, M$_{\star}<$10$^{10}$ galaxies make a negligible contribution to the XLF, in both MS and SB populations, out to z$\sim$2. Instead, at higher redshifts the galaxy M$_{\star}$ function steepens, meaning the number density of less massive to more massive systems increases, thus strenghtening the contribution of M$_{\star}<$10$^{10}$ galaxies, especially at faint L$_{\rm X}$.
As for galaxies at 10$^{10}<$M$_{\star}<$10$^{11}$~M$_{\odot}$, they tend to dominate the XLF up to the knee L$_{\rm X}$ of the corresponding population, thus contributing to the bulk X-ray emission at all redshifts. Lastly, galaxies with M$_{\star}>$10$^{11}$~M$_{\odot}$ contribute to the bright-end XLF at all cosmic epochs, both for MS and SB systems. Starbursts populate higher L$_{\rm X}$ than MS galaxies matched in M$_{\star}$ and redshift, given their $\lambda_{\rm EDD}$ distribution being shifted toward higher values.

\begin{figure}
     \includegraphics[width=\linewidth]{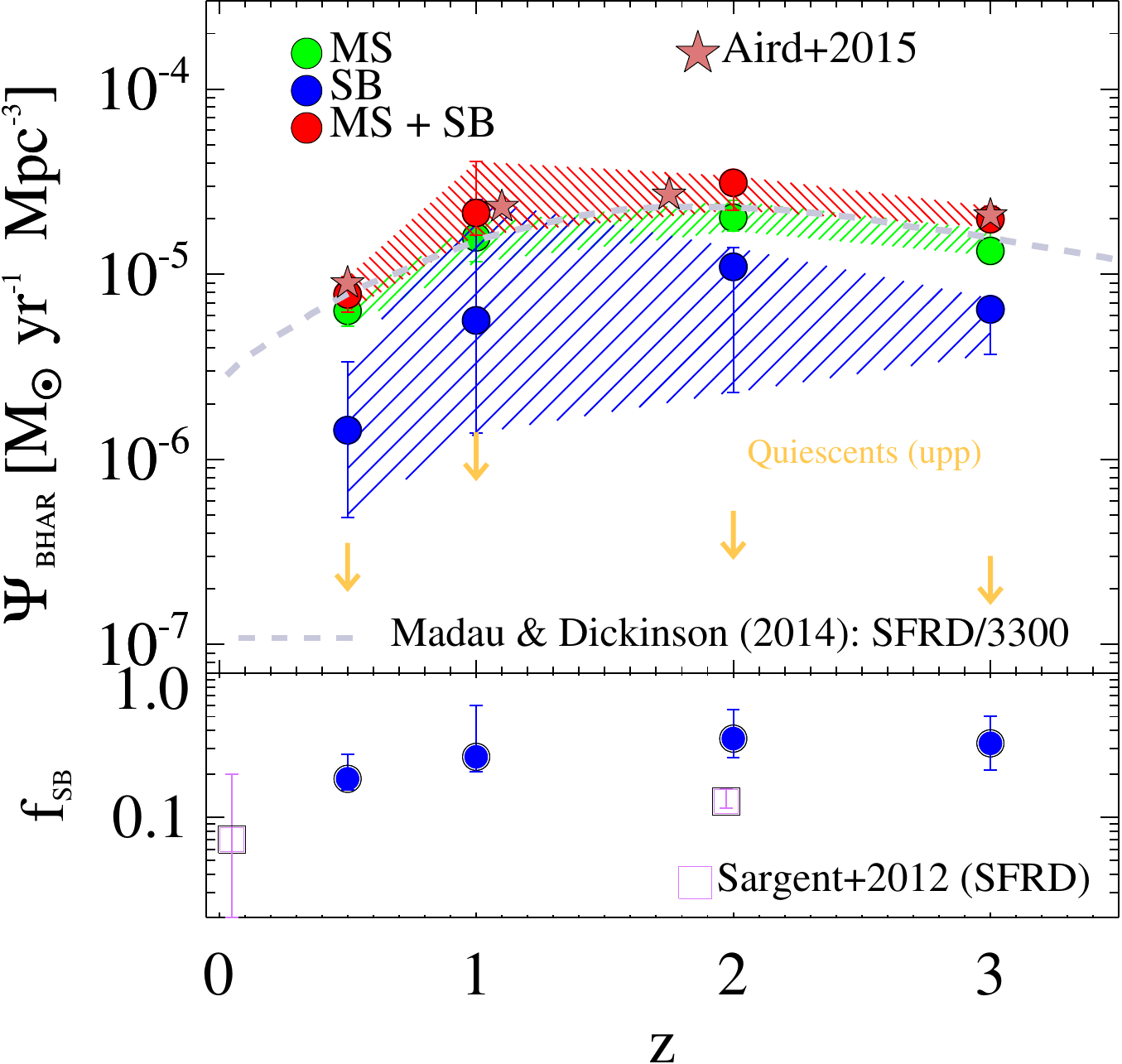}
 \caption{\small The SMBH accretion rate density $\Psi_{\rm BHAR}(z)$ since z$\sim$3 (red) split between MS (green) and SB (blue) galaxies. The MS population makes the bulk $\Psi_{\rm BHAR}$ at all cosmic epochs, while SBs are sub-dominant but evolve in a similar fashion. Downward arrows mark the upper limits on $\Psi_{\rm BHAR}$ obtained from quiescent galaxies (Section \ref{qxlf}). The overall BHARD agrees by design with the derivation by \citet{Aird+2015}, purple stars), and displays a similar shape to the star formation rate density (SFRD, \citealt{Madau+2014}), here scaled down by 3300 to ease the comparison (grey dashed line). As displayed in the bottom panel, we find f$_{\rm SB} \sim$20--30\% of the BHARD. Similarly, f$_{\rm SB} \sim$10--15\% of the full SFRD at z$\sim$2, and marginally consistent the local value (\citealt{Sargent+2012}, magenta squares). Error bars are provided at 1$\sigma$ level.
 }
   \label{fig:bhard}
\end{figure}

\subsection{Dissecting the SMBH accretion history among MS and SB galaxies}  \label{bhard}

Given the derived XLF, we are able to derive the Black Hole Accretion Rate Density (BHARD or $\Psi_{\rm BHAR}$) since z$\sim$3, separately between MS and SB galaxies. This quantity is fundamental for characterising the overall luminosity-weighted SMBH growth and is defined by the following expression:
\begin{equation}
\Psi_{\rm BHAR}(z) = \int_{0}^{\infty}{ \frac{1-\epsilon_{\rm rad}}{\epsilon_{\rm rad} ~ c^2} ~ L_{\rm AGN} ~ \phi(L_{\rm AGN}) ~ d\log L_{\rm AGN}}
\label{eq:bhard}
\end{equation}
where $\epsilon_{\rm rad}$ is the matter-to-radiation conversion efficiency, which is assumed for simplicity to take the constant value $\epsilon_{\rm rad}$=0.1, in line with previous studies (\citealt{Marconi+2004}; \citealt{Hopkins+2007}; \citealt{Merloni+2008}; \citealt{Delvecchio+2014}; \citealt{Ueda+2014}; \citealt{Aird+2015}). 

The AGN bolometric luminosity $L_{\rm AGN}$ is simply scaled from L$_{\rm X}$ via a set of k$_{\rm BOL}$. Once we remove the dependence on the single k$_{\rm BOL}$=22.4 used in previous studies (Sect.~\ref{bhsf}), we choose to adopt a set of L$_{\rm X}$--dependent k$_{\rm BOL}$ from \citet{Lusso+2012} when calculating our $\Psi_{\rm BHAR}(z)$ estimates.  Nevertheless, we will discuss below the effect that a constant k$_{\rm BOL}$=22.4 would have on the estimated $\Psi_{\rm BHAR}(z)$.

Fig.~\ref{fig:bhard} shows the $\Psi_{\rm BHAR}(z)$ derived from Eq.~\ref{eq:bhard} in our four redshift bins (red points). The relative contributions from MS and SB galaxies are the green and blue points, respectively. The crossing lines enclose the $\pm$1$\sigma$ uncertainties, by simply propagating the $\pm$1$\sigma$ confidence interval of the XLF (see Fig.~\ref{fig:xlf_aird}). We integrate the upper limit XLF of quiescent galaxies at each redshift (Section \ref{qxlf}), and report the corresponding upper limits on $\Psi_{\rm BHAR}(z)$ (orange downward arrows). As displayed in Fig. \ref{fig:bhard}, the MS population makes the bulk $\Psi_{\rm BHAR}$ at all cosmic epochs, while SBs are sub-dominant and display a similar redshift evolution. The integrated BHARD shown in Fig.~\ref{fig:bhard} agrees by design with the derivation by \citeauthor{Aird+2015} (\citeyear{Aird+2015}, purple stars), and displays a broadly similar shape to that of the star formation rate density (SFRD, \citealt{Madau+2014}), here scaled down by 3300 for illustrative purpose (grey dashed line). This similarity is a natural consequence of the redshift-invariant BHAR/SFR ratio constrained from our analysis (Section \ref{par_evo}), that is in turn a genuine result of our modeling. The bottom panel of Fig.~\ref{fig:bhard} displays the fractional contribution of the SB population (f$_{\rm SB}$) relative to the total $\Psi_{\rm BHAR}(z)$ at each redshift: f$_{\rm SB}$ ranges between 20 and 30\% and stays roughly constant with redshift. The redshift-invariant f$_{\rm SB}$ is, instead, artificially induced by our assumption that MS and SB galaxies share the same BHAR/SFR ratio, at each M$_{\star}$ and redshift. Indeed, the fraction of star formation rate density (SFRD) made by SB galaxies (\citealt{Sargent+2012}, magenta squares) ranges between 10 and 15\%, consistently with the two galaxy populations contributing in similar proportions to both SMBH accretion and star formation at all cosmic epochs (see also \citealt{Gruppioni+2013}; \citealt{Magnelli+2013}).

We note that adopting a L$_{\rm X}$--dependent k$_{\rm BOL}$ does change slightly the resulting BHARD relative to the case of a single k$_{\rm BOL}$. Indeed, at z=0.5 the assumption of k$_{\rm BOL}$=22.4 is consistent with the mean L$_{\rm X}$--evolving k$_{\rm BOL}$ at the break L$_{\rm X}$. Nevertheless, at z$>$0.5 the break L$_{\rm X}$ shifts to higher values, corresponding to about 2$\times$ higher k$_{\rm BOL}$, implying a differential boost of the integrated BHARD. Assuming a single k$_{\rm BOL}$=22.4 would instead lower f$_{\rm SB}$ down to $\approx$15--20\% at all redshifts.

Our choice of neglecting the quiescent galaxy population might lead us to slightly overestimate the relative contribution of MS and SB galaxies to the full BHARD. We quantify these fractions based on the derived upper limits on the quiescent $\Psi_{\rm BHAR}$, being: $<$4.3\% (z=0.5), $<$6.5\% (z=1), $<$1.8\% (z=2) and $<$1.6\% (z=3). If the total $\Psi_{\rm BHAR}$ was rescaled to accommodate the quiescent population, the relative f$_{\rm SB}$ would change by the following amount: from 20\% to 19\% at z=0.5; from 23\% to 21\% at z=1, and unchanged at higher redshifts. We note that these small differences represent upper limits (see Section \ref{xlf}). To this end, disentangling the small contribution of the quiescent population is beyond the scope of this paper (but see \citealt{Bernhard+2018}).

\begin{deluxetable*}{cccccccc||ccc}
\tablecaption{Uncertainties and confidence ranges (at 1$\sigma$ level) of the free input parameters assumed in this work ($\alpha$, $\beta$, $\lambda^{*}_{\rm MS}$, $\lambda^{*}_{\rm SB}$ and the BHAR/SFR relation with M$_{\star}$). In addition, we report the cross-over L$_{\rm X}$ (L$_{\rm X}^{\rm cross}$) and the fraction of $\lambda_{\rm EDD}$ distribution spent above 10\% Eddington (f[$\lambda_{\rm EDD}>$0.1]), for both MS and SB galaxies. The double vertical line separates free parameters (left) from other byproduct quantities (right).}
\tablewidth{0pt}
\tablehead{\colhead{Redshift} & \colhead{$\chi^2_{\rm red}$} & \colhead{$\alpha$} & \colhead{$\beta$} & \colhead{$\log \lambda^{*}_{\rm MS}$} &   \colhead{$\log \lambda^{*}_{\rm SB}$} & \colhead{slope} & \colhead{norm} & \colhead{L$_{\rm X}^{\rm cross}$}  & \colhead{f($\lambda_{\rm EDD, MS}>$0.1)} &   \colhead{f($\lambda_{\rm EDD, SB}>$0.1)} \\
    &   &   &   &   &  & BH/SF  & BH/SF &   &   &   }
\decimalcolnumbers
\startdata 
z=0.5  &   0.60   &  0.55$^{}_{}$             &   4$^{+1^{**}}_{-0}$  &  -0.7$^{+0.1}_{-0.2}$  &  0.0$^{+0.5^{**}}_{-0.5}$  &     0.95$^{+0.00}_{-0.30}$  &  -13.4$^{+3.1}_{-0.0}$   & 44.59$^{+0.09}_{-0.23}$    &  0.004$^{+0.001}_{-0.002}$  &   0.030$^{+0.006}_{-0.011}$    \\
z=1    &   0.29   &  0.22$^{+0.06}_{-0.00}$   &   5$^{+0^{**}}_{-2}$  &  -0.5$^{+0.1}_{-0.3}$  &  0.5$^{+0.0^{**}}_{-1.0}$  &     0.73$^{+0.22}_{-0.24}$  &  -11.2$^{+2.5}_{-2.2}$   & 44.74$^{+0.09}_{-0.26}$    &  0.016$^{+0.007}_{-0.004}$  &   0.038$^{+0.019}_{-0.021}$     \\
z=2    &   0.85   &  0.06$^{+0.05}_{-0.00}$   &   4$^{+0}_{-1}$  &  -0.4$^{+0.1}_{-0.1}$  &  0.4$^{+0.0}_{-0.8}$  &     0.73$^{+0.22}_{-0.29}$  &  -11.2$^{+3.0}_{-2.2}$   & 44.96$^{+0.09}_{-0.12}$    &  0.044$^{+0.006}_{-0.015}$  &   0.102$^{+0.001}_{-0.056}$     \\
z=3    &   1.32   &  0.03$^{+0.02}_{-0.00}$   &   5$^{+0^{**}}_{-0^{**}}$  &  -0.2$^{+0.2}_{-0.0}$  &  0.5$^{+0.0^{**}}_{-0.3}$  &     0.91$^{+0.00}_{-0.01}$  &  -13.0$^{+0.2}_{-0.0}$   & 45.15$^{+0.31}_{-0.15}$    &  0.065$^{+0.002}_{-0.001}$  &   0.153$^{+0.001}_{-0.041}$     \\
\hline    
\enddata
\tablecomments{Zero lower or upper errors are due to our discrete parameter grid, and they should be interpreted as smaller than the closest value. $(**)$ This symbol denotes those values touching the upper edge of our input grid, and should be taken as lower limits. }
\label{tab:uncertainties}
\end{deluxetable*}

\begin{figure}
     \includegraphics[width=\linewidth]{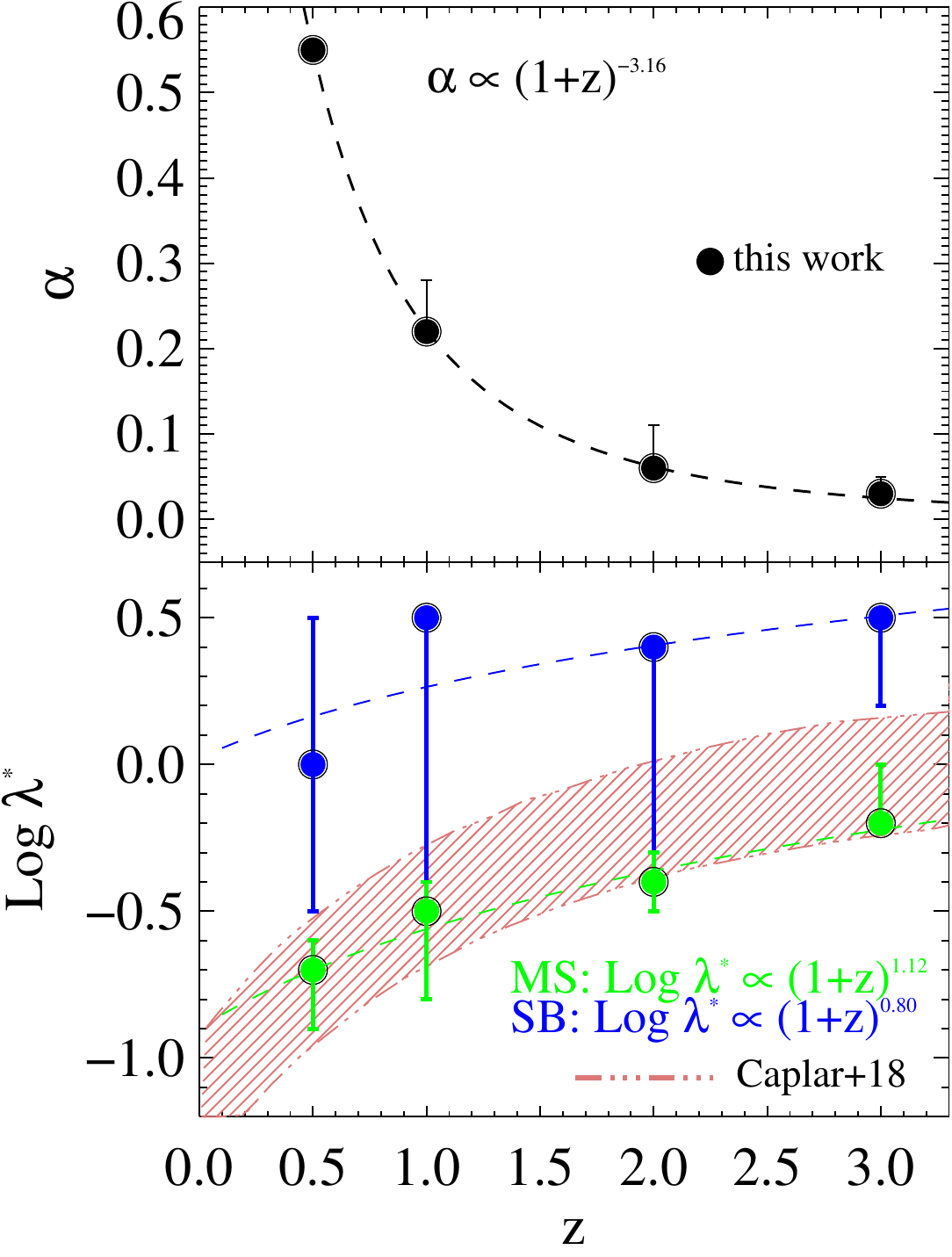}
 \caption{\small Top panel: redshift evolution of the faint-end slope of the $\lambda^{\rm EDD}$ ($\alpha$), in the form $\alpha \propto$(1+z)$^{\gamma}$. The best solution is given by $\gamma$=-3.16$^{+0.79}_{-0.00}$. Bottom panel: redshift evolution of the break $\lambda^{\rm EDD}$ for MS and SB galaxies (green and blue points, respectively). We fit a powerlaw trend finding $\lambda^{*}_{\rm MS} \propto$(1+z)$^{1.12}$ (MS) and $\lambda^{*}_{\rm SB} \propto$(1+z)$^{0.80}$ (SB). Uncertainties are given at 1$\sigma$ level. For comparison, the curve from \citet{Caplar+2018} is shown (red crossed area), valid for AGN host galaxies close to the knee of the galaxy M$_{\star}$ function and AGN bolometric LF, respectively. See Section~\ref{par_evo} for details.
 }
   \label{fig:par_evo}
\end{figure}

\subsection{Uncertainties and evolution of free parameters}  \label{par_evo}

In this Section we discuss the uncertainties on each free parameter assumed in this work, as well as their redshift evolution. As listed in Table \ref{tab:par}, five free parameters are adopted in this work: the faint and bright-end slopes of the $\lambda_{\rm EDD}$ distribution ($\alpha$, $\beta$), the break Eddington ratio of MS and SB galaxies ($\lambda^{*}_{\rm MS}$, $\lambda^{*}_{\rm SB}$) and the BHAR/SFR relation with M$_{\star}$. The combination of these parameters identifies the $\pm$1$\sigma$ confidence range around the best XLF shown in Fig.~\ref{fig:xlf_aird} (crossed lines). We checked that adding the SB component to the XLF improves the reduced $\chi^2$ at $>$90\% significance level in all redshift bins. 
Furthermore, from the inspection of the covariance matrices between all free parameters, we verified that their uncertainties seem to be unrelated to each other. Below we discuss the confidence range of each free parameter, and refer the reader to Table~\ref{tab:uncertainties} for a comprehensive list of uncertainties.

The faint-end slope $\alpha$ flattens with redshift in the form $\alpha \propto$(1+z)$^{\gamma}$ (Eq.\ref{eq:gamma}), which yields a best value of $\gamma$=-3.16$^{+0.79}_{-0.00}$. This implies a steep flattening (though not the steepest trend assumed in our input grid), corresponding to a nearly flat $\lambda_{\rm EDD}$ distribution at z=3 (top panel of Fig.~\ref{fig:par_evo}).

The bright-end slope $\beta$ is unconstrained from our analysis. Among the $\beta$ values initially allowed by our grid (2, 3, 4 and 5), the resulting $\pm$1$\sigma$ confidence intervals range between 3 and 5 (see Table \ref{tab:uncertainties}). Conversely, the flatter value of $\beta$=2 is marginally (at $>$1$\sigma$) disfavored, despite being the closest to what found in previous studies ($\beta \sim$1.8--2.5; e.g. \citealt{Hopkins+2007}; \citealt{Caplar+2015}, \citeyear{Caplar+2018}). This apparent discrepancy is likely driven by the fact that two $\lambda_{\rm EDD}$ distributions are here assumed to contribute to the bright-end XLF, with their relative breaks being placed at different L$_{\rm X}$. This overlap generates an overall flatter slope (e.g. $\beta \sim$2), that in our study is parametrised with two steeper slopes offset from each other.

The break $\lambda_{\rm EDD}$ of MS galaxies, $\lambda^{*}_{\rm MS}$, is let free to vary across the logarithmic range [-1; +0.5[, in all redshift bins. Our best solution prefers a progressive shift of $\lambda^{*}_{\rm MS}$ with redshift, from $\lambda^{*}_{\rm MS}$=-0.7 at z=0.5 to $\lambda^{*}_{\rm MS}$=-0.2 at z=3 (see Table \ref{tab:uncertainties}). We parametrise this redshift trend as follows (see green dashed line in Fig.~\ref{fig:par_evo}, bottom panel).
\begin{equation}
 \lambda^{*}_{\rm MS} =  10^{-0.9\pm 0.4} \cdot (1+z)^{1.12\pm0.19} 
 \label{eq:ledd_ms}
\end{equation}
Not only $\lambda^{*}_{\rm MS}$ is constrained fairly well by our analysis, but its evolution with redshift closely resembles the trend presented in \citeauthor{Caplar+2018} (\citeyear{Caplar+2018}, crossed area). The authors studied the characteristic $\lambda^{*}$ of AGN host galaxies based on the ratio between BHAR and M$_{\star}$ density (from \citealt{Aird+2015} and \citealt{Ilbert+2013}, respectively), from which they identified the corresponding knee L$_{\rm AGN}$ and M$_{\star}$. For the whole star-forming galaxy population (i.e. MS and SB), they obtained $\lambda^{*} \propto$(1+z)$^{2.5}$ out to z$\sim$2, and constant at z$>$2.

In addition, Fig.~\ref{fig:par_evo} (bottom panel) shows the evolution of the break $\lambda_{\rm EDD}$ of SB galaxies ($\lambda^{*}_{\rm SB}$, blue dashed line). 
\begin{equation}
 \lambda^{*}_{\rm SB} =  10^{0.0\pm 1.0} \cdot (1+z)^{0.80\pm0.57}
 \label{eq:ledd_sb}
\end{equation}
As previously mentioned, our prior assumption that MS and SB galaxies share the same BHAR/SFR ratio, at each M$_{\star}$ and redshift, produces a constant gap between $\lambda^{*}_{\rm SB}$ and $\lambda^{*}_{\rm MS}$ of the order of 0.8~dex (i.e. a factor of 6, Section \ref{bhard}). This gap mimics the $\times$6 higher SFR between SB and MS galaxies (\citealt{Schreiber+2015}; \citealt{Bethermin+2017}). Notwithstanding our prior assumptions induce the redshift trend of $\lambda^{*}_{\rm SB}$, empirical studies do support the condition $\lambda^{*}_{\rm SB} > \lambda^{*}_{\rm MS}$ (e.g. \citealt{Rodighiero+2015}; \citealt{Delvecchio+2015}; \citealt{Aird+2019}; \citealt{Grimmett+2019}; \citealt{Bernhard+2019}). An interesting implication is that SB galaxies are expected to host AGN accreting close or slightly above the Eddington limit, especially toward higher redshifts. This will be further discussed in Section~\ref{fgas}.

The BHAR/SFR vs. M$_{\star}$ relations assumed in this work include 18 trends taken from the recent literature, as displayed in Fig.~\ref{fig:bhar_sfr}. On the one hand, the best slope and normalisation listed in Table \ref{tab:uncertainties} at each redshift are consistent with a redshift-invariant relation. On the other hand, our analysis strongly supports a M$_{\star}$--dependent BHAR/SFR relation, with an average slope of 0.73$^{+0.22}_{-0.29}$, consistently with \citet{Aird+2019} within the uncertainties. We verified that flatter relations, like that proposed by \citet{Mullaney+2012} or the flat trend BHAR/SFR=10$^{-3}$, would lead to higher than observed faint-end XLF, due to excessive mean BHAR arising from low-mass galaxies.

\section{Discussion} \label{discussion}

In this Section we try to interpret our results in a broader context of AGN and galaxy evolution. Firstly, we validate our model predictions against known observed trends that have been reported in literature (Section~\ref{testing}). Next, we discuss the broad implications of our findings within a two-fold framework: (i) the role of cold gas content in driving SMBH accretion within MS and SB galaxies (Section~\ref{fgas}); (ii) the AGN duty cycle as a function of MS offset and redshift (Section~\ref{dutycycle}).

\subsection{Testing our modeling against the observed L$_{\rm X}$--SFR relation} \label{testing}

A number of studies have recently reported an apparently flat, or slightly positive, relationship between average SFR and L$_ {\rm X}$ for X-ray selected AGN at various redshifts. The origin of this trend is still debated. On the one hand, at moderate ($<$10$^{44}$~erg~s$^{-1}$) X-ray luminosities, a flat trend is commonly observed, which argues in favor of a weak dependence between SMBH accretion and star formation in galaxies \citep{Hickox+2014}, possibly driven by stochastic fuelling mechanisms that wash out a potential correlation with the instantaneous AGN accretion rate traced by X-ray emission (e.g. \citealt{Rosario+2012}; \citealt{Stanley+2015}).

On the other hand, at the highest (quasar-like, $>$10$^{44}$~erg~s$^{-1}$) X-ray luminosities, other studies argue in favor of a slightly positive trend (e.g. \citealt{Netzer+2016}; \citealt{Duras+2017}; \citealt{Schulze+2019}), suggesting concomitant star formation and AGN activity, possibly driven by major mergers. The transition between these two modes is still poorly understood, as it depends not only on L$_{\rm X}$, but also on sample selection and redshift. Testing our modeling with a number of observed L$_{\rm X}$--SFR trends is therefore a useful test case for double-checking the validity of our approach based on solid empirical grounds. 

\begin{figure}
     \includegraphics[width=\linewidth]{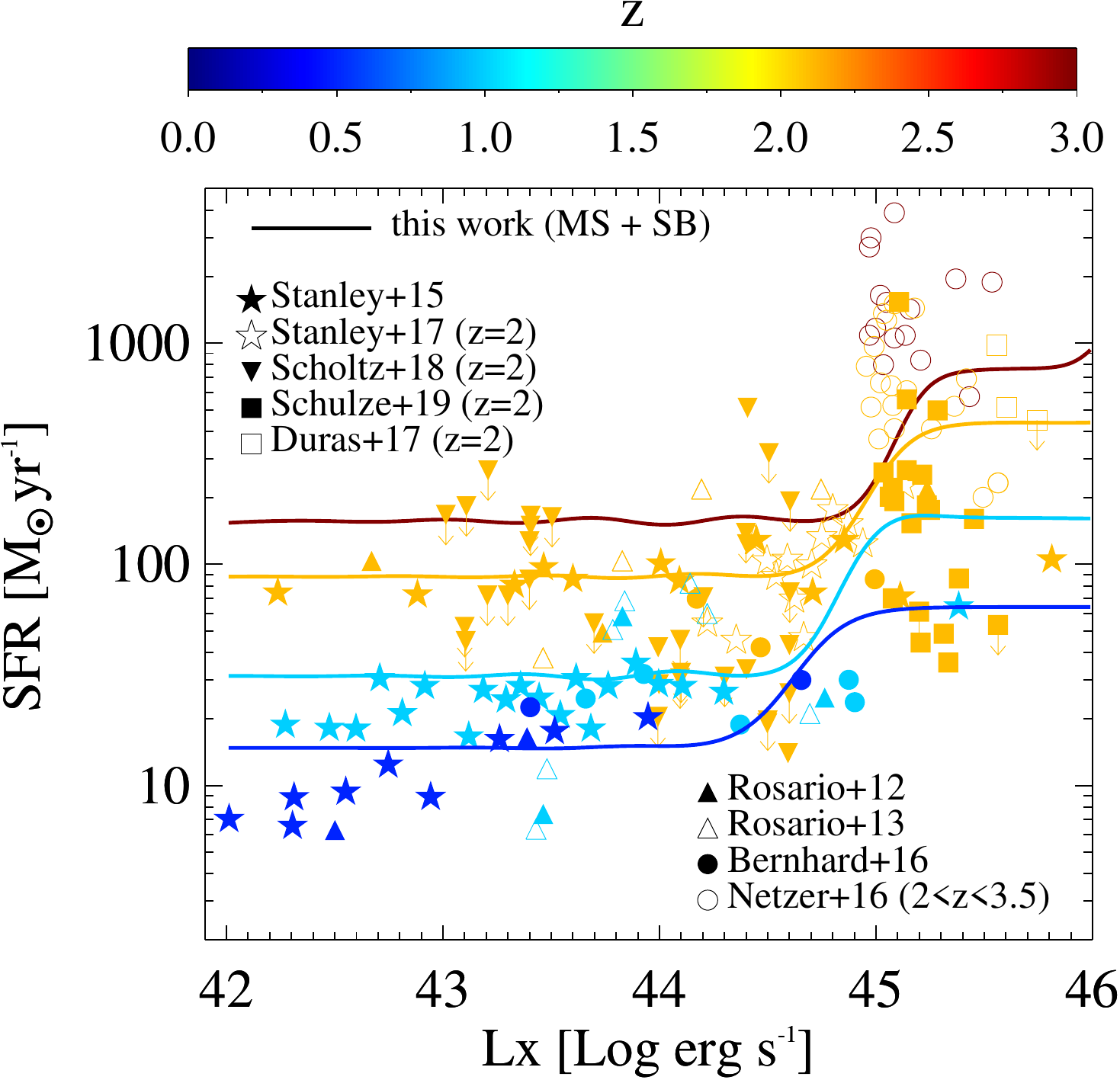}
 \caption{\small Linear mean SFR derived in bins of L$_{\rm X}$ and redshift, by weighing the total (MS and SB combined) XLF by the SFR of each galaxy population contributing at each L$_{\rm X}$. Solid lines represent our predicted trends, while datapoints are from observational samples of X-ray selected AGN. The colorbar indicates the redshift of each dataset. We note that the SFR of each sample was properly scaled to match the closest redshift assumed in this work, by a factor corresponding to the evolution of the MS normalisation between the two redshifts (at M$_{\star}$=10$^{10.8}$~M$_{\odot}$). Our modeling shows a good agreement with the observed L$_{\rm X}$--SFR trends at all redshifts. The ``bump'' predicted by our curves at high L$_{\rm X}$ is driven by the gradual predominance of SB galaxies, with the position of the bump scaling with redshift (see Fig.~\ref{fig:frac_SB}).  
 }
   \label{fig:flat_test}
\end{figure}

We collect a compilation of L$_{\rm X}$--SFR trends across our full redshift range 0.5$<$z$<$3, as shown in Fig.~\ref{fig:flat_test}. Samples were taken from \citeauthor{Stanley+2015} (\citeyear{Stanley+2015}, filled stars); \citeauthor{Rosario+2012} (\citeyear{Rosario+2012}, filled triangles); \citeauthor{Rosario+2013} (\citeyear{Rosario+2013}, empty triangles); \citeauthor{Bernhard+2016} (\citeyear{Bernhard+2016}, filled circles). At z$\sim$2 we collect data from \citeauthor{Stanley+2017} (\citeyear{Stanley+2017}, empty stars); \citealt{Scholtz+2018} (\citeyear{Scholtz+2018}, downward triangles); \citeauthor{Duras+2017} (\citeyear{Duras+2017}, empty squares); \citeauthor{Schulze+2019} (\citeyear{Schulze+2019}, filled squares) and \citeauthor{Netzer+2016} (\citeyear{Netzer+2016}, empty circles), this latter extending out to z$\sim$3.5. The colorbar indicates the redshift of each dataset.

Solid lines highlight our linear mean SFR estimates, derived in bins of L$_{\rm X}$ and redshift, by weighing the total (MS and SB combined) XLF by the SFR of each galaxy population contributing at each L$_{\rm X}$. We note that each dataset was properly re-scaled in SFR to match the closest redshift assumed in this work, by a factor corresponding to the evolution of the MS normalisation between the two redshifts (at M$_{\star}$=10$^{10.8}$~M$_{\odot}$). Whenever necessary, we converted the SFRs taken from the literature to a \citet{Chabrier2003} IMF. Our modeling displays a good agreement with the observed L$_{\rm X}$--SFR trends at all redshifts. The high-L$_{\rm X}$ ``bump'' predicted by our curves is likely driven by the gradual predominance of SB galaxies, with the position of the bump shifting with redshift (see top panel of Fig.~\ref{fig:par_evo2}).  

As discussed in \citet{Stanley+2015}, the predicted mean SFR are strongly dependent on the assumed $\lambda_{\rm EDD}$ distribution, that incorporates the stochasticity of SMBH accretion. However, we are able to circumvent this issue by constraining the main $\lambda_{\rm EDD}$ distribution slope and break via empirical BHAR/SFR relations and comparison with the observed XLF. This test is encouranging, as it demonstrates that our simple, empirically-motivated modeling, is successful in predicting the average SFR observed across a wide range of L$_{\rm X}$ and redshift.

\begin{figure}
     \includegraphics[width=\linewidth]{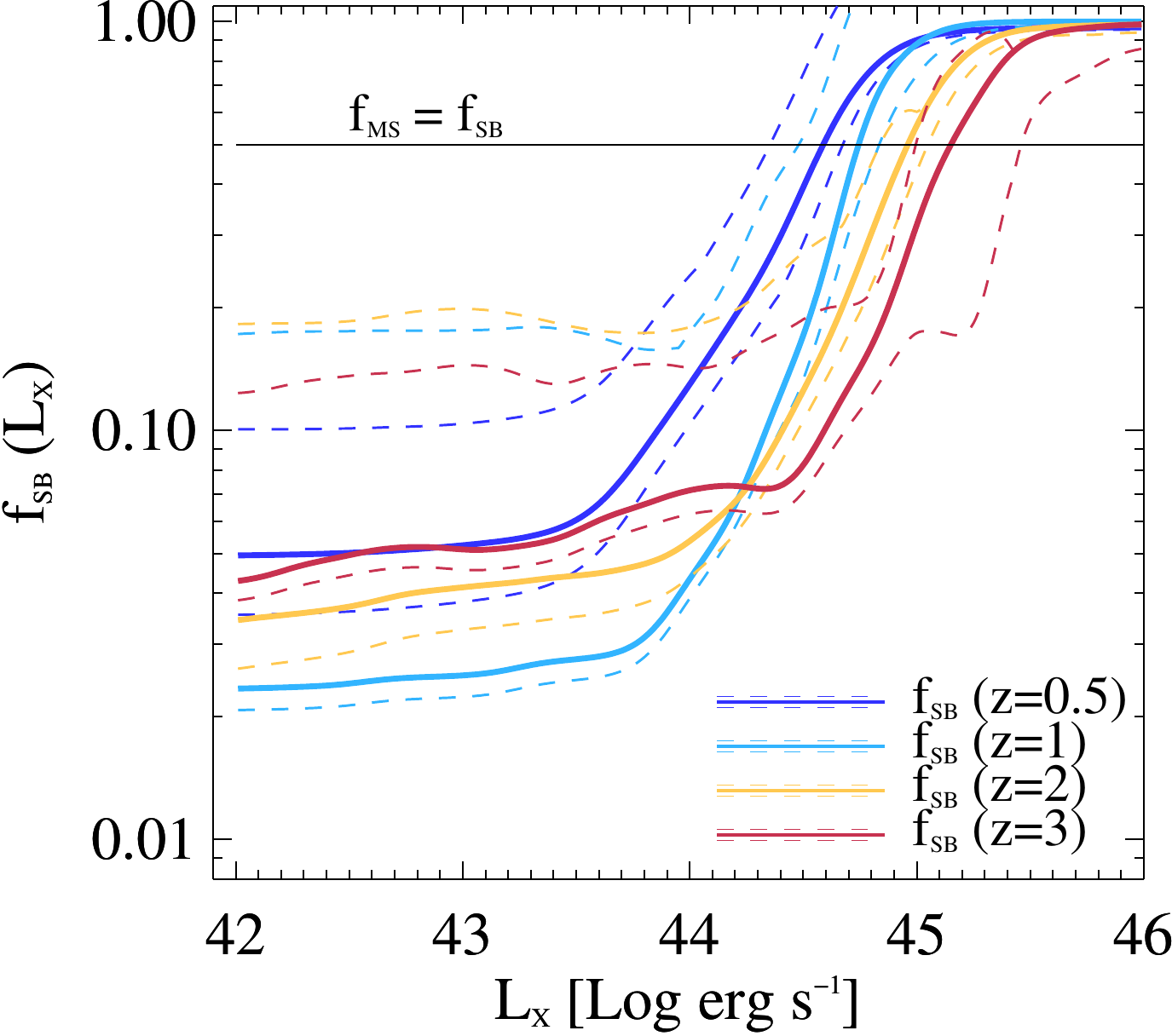}
 \caption{\small Ratio between SB-related XLF and total (MS and SB) XLF, namely f$_{\rm SB}$(L$_{\rm X}$), as a function of L$_{\rm X}$ (solid lines) and redshift (different colors). Dashed lines indicate the corresponding $\pm$1$\sigma$ confidence XLF propagated to f$_{\rm SB}$. While the SB fraction accounts for just few~\% at L$_{\rm X}<$10$^{43.5}$~erg~s$^{-1}$, it increases to nearly 100\% at L$_{\rm X}>$10$^{45}$~erg~s$^{-1}$, due to the gradual predominance of SB galaxies at the highest L$_{\rm X}$. The black horizontal line marks f$_{\rm SB}$=0.5, which identifies the cross-over L$_{\rm X}^{\rm cross}$. 
 }
   \label{fig:frac_SB}
\end{figure}

\subsection{What drives the evolving SMBH growth in MS and SB galaxies?} \label{fgas}

\begin{figure}
     \includegraphics[width=\linewidth]{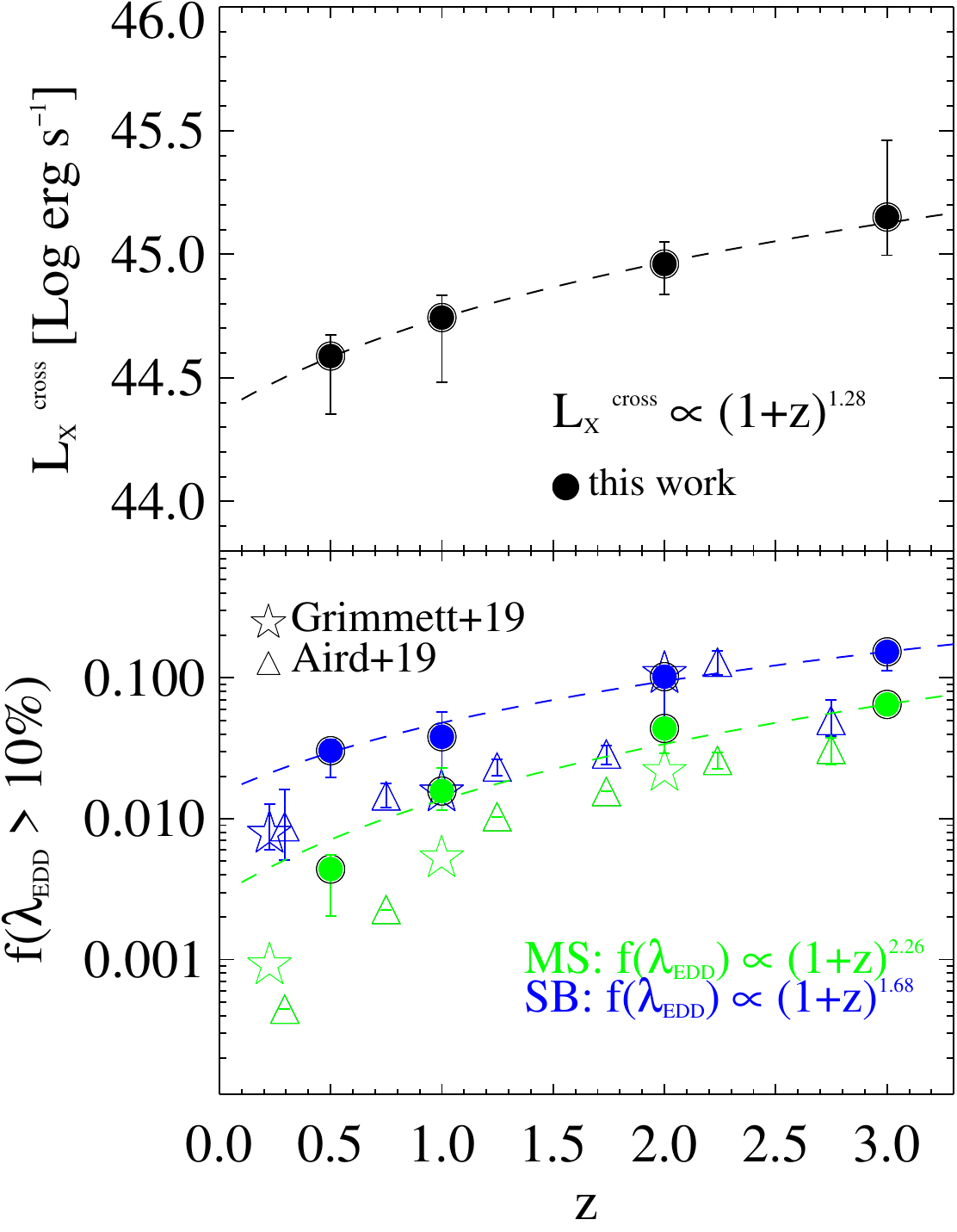}
 \caption{\small Top panel: cross-over L$_{\rm X}$ (L$_{\rm X}^{\rm cross}$) as a function of redshift. The dashed line corresponds to a powerlaw fit as L$_{\rm X}^{\rm cross} \propto$(1+z)$^{1.28}$, suggesting a roughly linear increase. Bottom panel: fraction of the $\lambda_{\rm EDD}$ distribution above 10\% Eddington (f[$\lambda_{\rm EDD}>$0.1]), for both MS (green points) and SB (blue points) galaxies. Error bars are given at 1$\sigma$ level. Dashed lines indicate the powerlaw fit of both populations. For comparison, data taken from \citet{Grimmett+2019} and \citet{Aird+2019} are reported, based on more complex $\lambda_{\rm EDD}$ functions. Though our f[$\lambda_{\rm EDD}>$0.1] estimates stand sligthly higher than their data, we consistently observe a monotonic increase with redshift, with SB galaxies displaying higher mean $\lambda_{\rm EDD}$ (see Section~\ref{dutycycle} for details).    
 }
   \label{fig:par_evo2}
\end{figure}

Our study supports a not negligible contribution of SB galaxies to the integrated BHARD (20--30\%, see Fig.~\ref{fig:bhard}). While the bulk SMBH accretion history is made by massive (10$^{10}<$M$_{\star}<$10$^{11}$~M$_{\odot}$) MS galaxies, the SB population appears to take over at relatively high L$_{\rm X}$, enabling us to reproduce the bright-end XLF since z$\sim$3. Specifically, Fig.~\ref{fig:frac_SB} displays the fractional contribution of SB galaxies to the XLF (f$_{\rm SB}$), as a function of L$_{\rm X}$ and redshift. Solid lines mark the best f$_{\rm SB}$(L$_{\rm X}$), with colors used to indicate our four redshift values. Dashed lines indicate the corresponding $\pm$1$\sigma$ interval propagated from the XLF of SBs. We observe that f$_{\rm SB}$ increases with L$_{\rm X}$ from few~\% (L$_{\rm X}<$10$^{43.5}$~erg~s$^{-1}$) to nearly 100\% (L$_{\rm X}>$10$^{45}$~erg~s$^{-1}$). As expected, the scatter becomes narrower toward higher L$_{\rm X}$, at which SBs increasingly dominate. The black horizontal line marks f$_{\rm SB}$=0.5, which identifies the cross-over L$_{\rm X}^{\rm cross}$ (see top panel of Fig. \ref{fig:par_evo2}). This value scales as:
\begin{equation}
L_{\rm X}^{\rm cross} [\rm erg~s^{-1}] = 10^{44.36\pm0.20} \cdot(1+z)^{1.28 \pm 0.33}
\label{eq:lx_cross} 
\end{equation}

Interestingly, also the total IR (rest-frame 8--1000~$\mu$m) galaxy LF derived with deep \textit{Herschel} data displays a strong luminosity (and density) evolution with redshift, as well as a constant f$_{\rm SB} \sim$20\% (\citealt{Gruppioni+2013}; \citealt{Magnelli+2013}). This similarity is consistent with star formation and AGN accretion in galaxies evolving in a similar fashion over cosmic time, and similarly between MS and SB galaxies.

From our analysis, the best BHAR/SFR slope with M$_{\star}$ is found to range between 0.73 and 0.95, without a significance redshift dependence (see Table \ref{tab:uncertainties}). These values agree with the linear fit obtained by \citet{Aird+2019}, while they seem to reject at $\sim$3$\sigma$ significance a flat BHAR/SFR trend with M$_{\star}$. This positive M$_{\star}$ dependence introduces a non-linearity in the cosmic buildup of galaxies and their central SMBHs: while at low M$_{\star}$ the galaxy grows in mass faster than the SMBH (i.e. low BHAR/SFR ratio), when the galaxy reaches high enough M$_{\star}$ the SMBH gradually catches up (high BHAR/SFR ratio). This twofold behavior might be primarily driven by the ability of the dark matter halo mass in setting the usable amount of cold gas for the host \citep{Delvecchio+2019}.

As mentioned in Section \ref{par_evo}, another genuine trend is the observed shift of the $\lambda_{\rm EDD}$ distribution with redshift, which comes directly from the comparison with the observed XLF of \citet{Aird+2015}. As a consequence, our results suggest that SB galaxies have a characteristic $\lambda^{*}$ close or slightly above the Eddington limit (Section~\ref{par_evo}). Although this might sound unlikely, we note that (i) the uncertainties are broadly consistent with the Eddington limit; (ii) the Eddington limit is not a physical boundary, but it can be exceeded in case of non-spherical gas accretion; (iii) theoretical predictions of the $\lambda_{\rm EDD}$ distribution support a progressive flattening at low $\lambda_{\rm EDD}$, as well as an increasing fraction of super-Eddington accretion with redshift (e.g. \citealt{Kawaguchi+2004}; \citealt{Shirakata+2019}); (iv) we verified that imposing a maximum $\lambda_{\rm EDD}$ equal to the Eddington limit fails to reproduce the bright-end XLF at z$>$1, thus supporting the need for a minor fraction (1--5\%) of super-Eddington accretion in SB (while only $<$0.1\% for MS) galaxies at $>$3$\sigma$ significance.

Our results reveal a significant evolution of $\lambda^{*}_{\rm MS}$ with redshift, which induces the offset trend for SB galaxies (bottom panel of Fig. \ref{fig:par_evo}). This close-to-linear redshift dependence suggests that the probability of finding SMBHs accreting above a certain $\lambda_{\rm EDD}$ is higher toward earlier times, for both populations. A qualitatively similar cosmic evolution of the active AGN fraction and $\lambda_{\rm EDD}$ distribution is also independently seen in optically-selected quasars at 1$<$z$<$2 \citep{Schulze+2015} and ascribed to an increasing intensity of SMBH growth toward earlier times (see their Figs. 18 and 23).

It is well established that galaxies were more gas rich at earlier epochs, with the cold gas fraction f$_{\rm gas}$ increasing out to z$\sim$2--3 (as f$_{\rm gas}\propto$(1+z)$^2$, e.g. \citealt{Saintonge+2013}). Larger (molecular) gas reservoirs coincide with higher SFR densities via the SK relation, but also with more probable triggering of the central SMBH and the onset of radiative AGN activity (e.g. \citealt{Alexander+2012}, \citealt{Vito+2014}). In addition, higher redshift galaxies tend to be more compact and to show more disturbed and irregular morphologies (\citealt{Forster-Schreiber+2009}; \citealt{Kocevski+2012}). This profound structural transformation of galaxies over cosmic time might explain the concomitant evolution of the typical $\lambda_{\rm EDD}$ of their central SMBHs. A good place to witness the enhancement of both phenomena is given by high-redshift (z$>$1) SB galaxies, which are characterised by higher SFE and denser gas reservoirs (\citealt{Daddi+2010b}; \citealt{Genzel+2010}) relative to MS analogs at the same redshift. In these systems, several mechanisms, such as major mergers, violent disk instabilities \citep{Bournaud+2011}, cold gas inflows \citep{DiMatteo+2012} might be at play, triggering starbursting star formation, mostly enshrouded in compact and highly obscured molecular clouds. This is observed in numerical simulations to be the case at high redshift, where the typical M$_{\rm gas}$ exceeds M$_{\star}$ (\citealt{Dubois+2014}, \citeyear{Dubois+2016}), which makes them ideal environments for triggering highly-accreting SMBHs ($\lambda_{\rm EDD}>$10\%).  

In the light of the above considerations, our results support a picture in which the cosmic evolution of galaxies' cold gas content might be the main driver of the redshift-invariant BHAR/SFR relation and the positive shift of the $\lambda_{\rm EDD}$ distribution.

\subsection{The AGN duty cycle in MS and SB galaxies} \label{dutycycle}

The evolution of the characteristic $\lambda^{*}$ with redshift in both MS and SB galaxies links to the question of whether the AGN duty cycle changes over cosmic time. Putting constraints on the relative time spent by AGN above a certain $\lambda_{\rm EDD}$ is crucial for understanding the global incidence of AGN activity across the galaxy population. 

Following previous studies on this topic (\citealt{Aird+2019}; \citealt{Grimmett+2019}) we explore the fraction of AGN accreting above 10\% Eddington (f[$\lambda_{\rm EDD}>$0.1]) and how this evolves with redshift across MS and SB galaxies. We consider the $\lambda_{\rm EDD}$ distribution corresponding to the best XLF at each redshift and M$_{\star}$, separately for MS and SB, and integrate each function from $\lambda_{\rm MAX}$=100 (our maximum value) down to $\lambda_{\rm EDD}$=0.1. Because the $\lambda_{\rm EDD}$ is normalised to unity (Eq.~\ref{eq:norm}), it describes the stochasticity of SMBHs across the entire galaxy lifecycle. Therefore we intepret the fraction above a certain $\lambda_{\rm EDD}$ as proportional to the time spent above that $\lambda_{\rm EDD}$. At each redshift, we weight all the f[$\lambda_{\rm EDD}>$0.1] estimates obtained at various M$_{\star}$ by the contribution of each M$_{\star}$-bin to the total BHARD at that redshift (Section \ref{xlf_mass}). This way we infer the luminosity-weighted f[$\lambda_{\rm EDD}>$0.1] at each redshift, separately for MS and SB galaxies.
The bottom panel of Fig.~\ref{fig:par_evo2} shows f[$\lambda_{\rm EDD}>$0.1] for both MS (green points) and SB (blue points) galaxies. Error bars are given at 1$\sigma$ level, and incorporate the propagation of the XLF uncertainties. We fit the trend for MS and SB galaxies with a powerlaw function in (1+z), obtaining the following expressions:
\begin{equation}
 f[\lambda_{\rm EDD, MS}>0.1] =  10^{-2.54 \pm 0.19} \cdot (1+z)^{2.26 \pm 0.40}
 \label{eq:frac_10_ms}
\end{equation}
\begin{equation}
 f[\lambda_{\rm EDD, SB}>0.1] =  10^{-1.82 \pm 0.57} \cdot (1+z)^{1.68 \pm 1.02}
 \label{eq:frac_10_sb} 
\end{equation}
which imply a steep rising toward earlier cosmic epochs, from 0.4\% (3.0\%) at z=0.5 to 6.5\% (15.3\%) at z=3 in MS (SB) galaxies.
Data taken from \citeauthor{Grimmett+2019} (\citeyear{Grimmett+2019}, stars) and \citeauthor{Aird+2019} (\citeyear{Aird+2019}, triangles) are reported for comparison, based on more complex $\lambda_{\rm EDD}$ profiles. Though our f[$\lambda_{\rm EDD}>$0.1] estimates stand sligthly higher than their data, we consistently observe a monotonic increase with redshift, suggesting that SMBHs spend longer at high accretion rates at earlier epochs, with SMBHs in SB galaxies being the most active. 

We stress that a \textit{longer} AGN duty cycle does not necessarily imply a longer duration of single episodes of AGN activity, but also similar timescales of AGN activity repeated more frequently across the galaxy lifecycle. Therefore, the AGN duty cycle we refer to coincides with the \textit{average} fraction of the SMBH lifetime spent above a given $\lambda_{\rm EDD}$.

Interestingly, our trend broadly follows the evolution of the molecular gas fraction f$_{\rm gas}$ with redshift, namely f$_{\rm gas}\propto$(1+z)$^{2}$ (e.g. \citealt{Saintonge+2013}; \citealt{Tacconi+2013}). We speculate that a link between f$_{\rm gas}$ and typical AGN Eddington ratio ($\propto$L$_{\rm X}$/M$_{\rm BH}$ $\sim$L$_{\rm X}$/M$_{\star}$ if assuming a fixed M$_{\rm BH}$--M$_{\star}$ ratio) could be foreseen if the available cold gas supply regulates the triggering and duration of both stellar and BH growth. At higher redshift, larger gas reservoirs are condensed in more compact regions than in local galaxies, so the gas depletion timescale (t$_{\rm dep}$=M$_{\rm gas}$/SFR) is shorter and star formation takes place more likely and more efficiently (\citealt{Daddi+2010b}; \citealt{Genzel+2010}). Moreover, in the early Universe the merger rate was significantly higher than today \citep{LeFevre+2013}, providing a further mechanism for fueling and sustaining SMBH-galaxy growth. In this scenario, if SB-driven star formation and BH accretion are mostly triggered by major mergers (e.g. \citealt{Calabro+2019}; \citealt{Cibinel+2019}), it might be plausible to expect higher average AGN accretion rates and longer AGN duty cycle (e.g. \citealt{Glikman+2015}; \citealt{Ricci+2017}), as we observe in this work.

\begin{figure}
     \includegraphics[width=\linewidth]{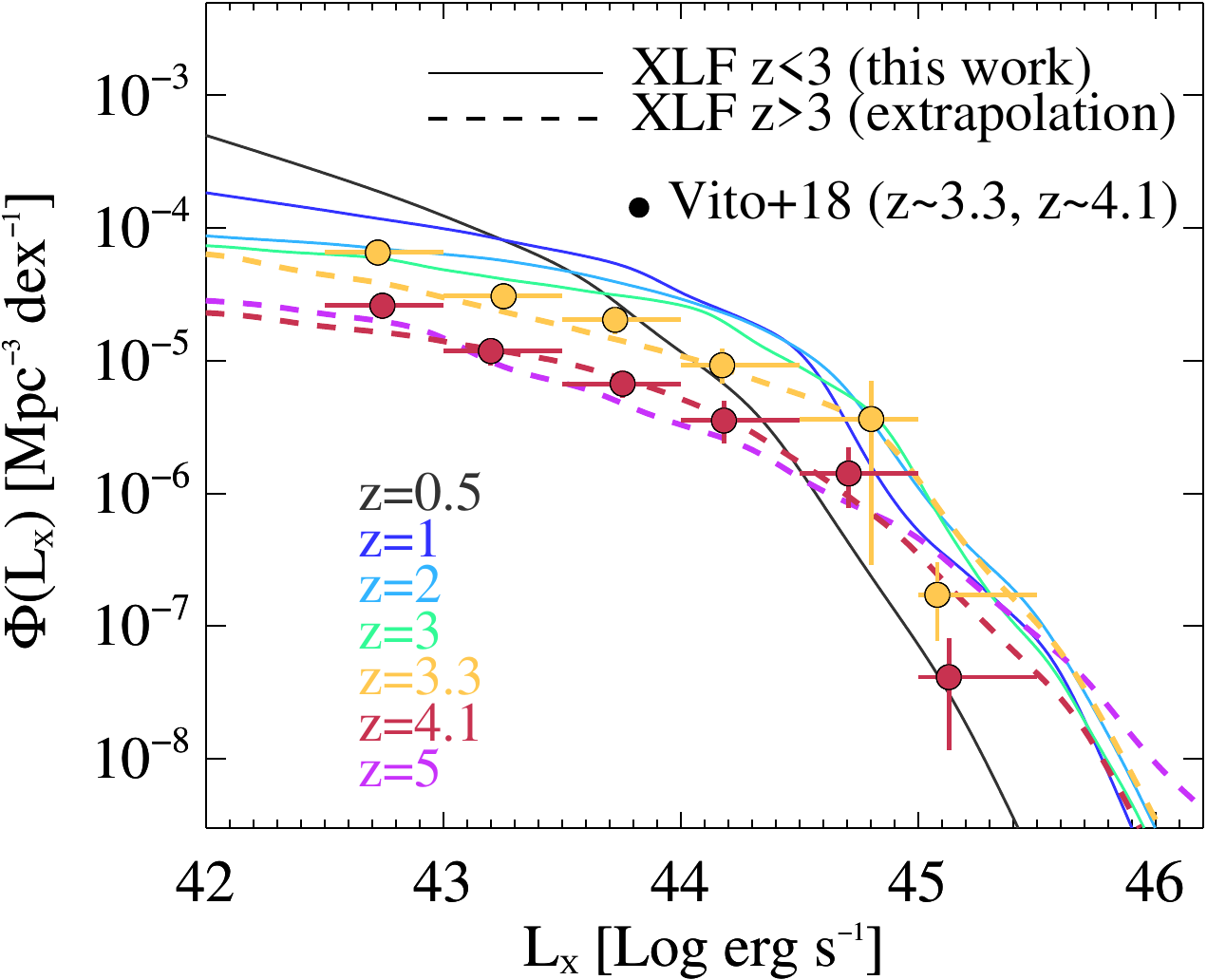}
 \caption{\small Best predicted XLFs at 0.5$<$z$<$3 (solid lines), and their extrapolation out to z$\sim$5 (dashed lines). Observational data from \citet{Vito+2018} are shown for comparison at z$\sim$3.3 (yellow circles) and z$\sim$4.1 (red circles). See text for details.
 }
   \label{fig:extrapolation}
\end{figure}
%
%
\subsection{Extrapolating the XLF out to z$\sim$5} \label{extrapolation}

Recent studies (e.g. \citealt{Vito+2018}; \citealt{Cowie+2020}) tried to constrain the XLF at 3$<$z$<$6 based on the currently deepest \textit{Chandra} data in the 7~Ms \textit{Chandra} Deep Field South \citep{Luo+2017}. An interesting outcome of those observational studies is that the global BHARD seems to display a steep decline at z$>$3, much stronger than that observed at z$<$1. Since these findings link directly to the integral of the XLF, here we compare our modeling against the most recent data at z$>$3, to further test whether our extrapolated XLF at z$>$3 yields a similar behavior (Fig.~\ref{fig:extrapolation}). To this end, we collect the latest observed XLF at 3$<$z$<$6 from \citet{Vito+2018}, centered at z$\sim$3.3 (3$<$z$<$3.6, yellow circles) and z$\sim$4.1 (3.6$<$z$<$6, red circles). Then we extrapolate our best XLF (obtained at 0.5$<$z$<$3), as follows. We convolve the galaxy M$_{\star}$ function \citep{Davidzon+2017} of MS and SB galaxies, at z$\sim$3.3 and z$\sim$4.1, with the corresponding $\lambda_{\rm EDD}$ distributions. These latter are derived by extrapolating the redshift evolution of the slopes ($\alpha$, $\beta$) and break $\lambda^{*}$ of each corresponding population (see Table~\ref{tab:uncertainties} and Eqs.~\ref{eq:ledd_ms}-\ref{eq:ledd_sb}). As shown in Fig.~\ref{fig:extrapolation}, our predicted XLF shows a very good agreement with current observational data \citep{Vito+2018} at z$\sim$3.3 and z$\sim$4.1, which we think further proves our modeling solid. We note that the CDF-S is a pencil-beam field, therefore the brightest data points of \citet{Vito+2018} are more uncertain.

As discussed in Sect.~\ref{eddratio}, the behavior of the XLF out to z$\sim$5 is driven by the a progressive flattening of the faint-end and a positive shift of the bright-end $\lambda_{\rm EDD}$ distribution. As confirmed by previous studies (e.g. \citealt{Aird+2012}; \citealt{Weigel+2017}; \citealt{Caplar+2018}), such a trend is consistent with an anti-hierarchical growth of BHs, possibly linked to an intrinsically longer AGN duty cycle (Sect.~\ref{dutycycle}).

We publicly release our best XLF (and its $\pm$1$\sigma$ confidence intervals) out to z$\sim$5 in the online supplementary material, for both MS and SB galaxies.

\subsection{From sBHAR to intrinsic Eddington ratio: super-Eddington growth?} \label{ledd_true}

We briefly explore the implications of relaxing the assumption of a constant M$_{\rm BH}$/M$_{\star}$ = 1/500 (e.g. \citealt{Haring+2004}, see Fig. \ref{fig:ledd_plot}). We stress again that such assumption does not enter our modeling, which is fully based on the \textit{observed} L$_{\rm X}$/M$_{\star}$ (i.e. sBHAR), while it comes into play when conceptually linking sBHAR to Eddington ratio (e.g. \citealt{Aird+2012}). In this respect, recent studies support a M$_{\star}$--increasing M$_{\rm BH}$/M$_{\star}$ ratio in star-forming galaxies, from both observational (\citealt{Reines+2015}; \citealt{Shankar+2016}; \citealt{Shankar+2019}) and theoretical (e.g. \citealt{Habouzit+2017}; \citealt{Bower+2017}; \citealt{Lupi+2019}) arguments. An intriguing implication of this behavior is presented in \citet{Delvecchio+2019}, where we integrate over time the BHAR/SFR trend obtained in this work (Section \ref{par_evo}) to track the cosmic assembly of the M$_{\rm BH}$--M$_{\star}$ scaling relation. In agreement with the above-mentioned literature, we found that in galaxies with M$_{\star}\gtrsim$10$^{10}$~M$_{\odot}$ the M$_{\rm BH}$/M$_{\star}$ ratio increases with M$_{\star}$ as: $\log$(M$_{\rm BH}$/M$_{\star}$) = -11.14 + 0.70$\cdot\log$M$_{\star}$. In less massive galaxies, instead, the relation may flatten out depending on the assumed BH seed mass, which typically ranges between 10$^2$ and 10$^6$ M$_{\odot}$ (e.g. \citealt{Begelman+1978}). 

\begin{figure}
     \includegraphics[width=\linewidth]{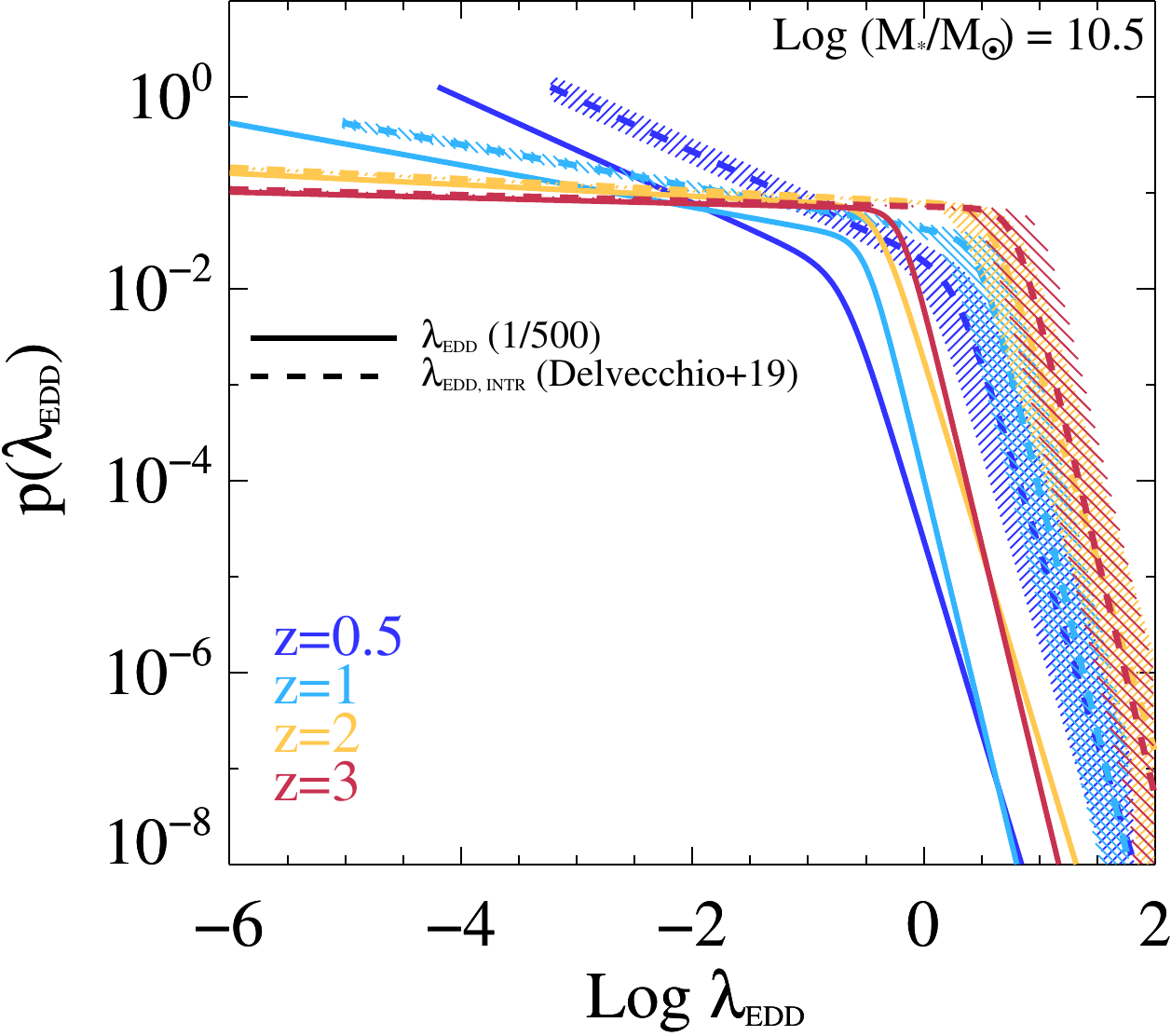}
 \caption{\small Comparison between $\lambda_{\rm EDD}$ distributions of MS galaxies at M$_{\star}$=10$^{10.5}$~M$_{\odot}$. Solid lines mark the p($\log \lambda_{\rm EDD}$) used in this work, based on the assumption that M$_{\rm BH}$/M$_{\star}$ = 1/500 (e.g. \citealt{Haring+2004}, see Fig. \ref{fig:ledd_plot}). Dashed lines show the equivalent distribution shifted in the \textit{intrinsic} $\lambda_{\rm EDD}$ space by assuming a M$_{\star}$--dependent M$_{\rm BH}$/M$_{\star}$ ratio for typical MS galaxies \citep{Delvecchio+2019}, incorporating the scatter of the M$_{\rm BH}$--M$_{\star}$ conversion (shaded area). Colours mark our four redshift bins. 
 }
   \label{fig:ledd_true}
\end{figure}

In this work, we have shown that the bulk of SMBH growth occurs in MS galaxies with 10$^{10}<$M$_{\star}<$10$^{11}$~M$_{\odot}$ (Section \ref{xlf_mass} and Fig. \ref{fig:xlf_aird_mass}). In that range, the relation obtained in \citet{Delvecchio+2019} yields M$_{\rm BH}$/M$_{\star}\approx$1/5000. From Eq.~\ref{eq:eddratio}, this implies that the \textit{intrinsic} Eddington ratio ($\lambda_{\rm EDD, INTR}$) would shift up by a factor of ten. Fig. \ref{fig:ledd_true} shows the comparison between the standard $\lambda_{\rm EDD}$ distributions (M$_{\rm BH}$/M$_{\star}$ = 1/500, solid lines) obtained in this work against the $\lambda_{\rm EDD, INTR}$ distributions (dashed lines), for MS galaxies at different redshifts. The effect of adopting empirical M$_{\rm BH}$/M$_{\star}$ ratios is that the break $\lambda_{\rm EDD, INTR}$ shifts \textit{above} the Eddington limit. This trend is significant, given the relatively small uncertainties on our break $\lambda_{\rm EDD}$ (see Table~\ref{tab:uncertainties}). Although we remind the reader that the Eddington limit is physically binding only for idealised conditions of BH accretion (see Section \ref{fgas}), our framework predicts between 0.5\% (z=0.5) and 5\% (z=3) super-Eddington BH growth in massive MS galaxies. The mean $\lambda_{\rm EDD, INTR}$ rises from 0.03 (z=0.5) to 0.19 (z=3), thus well below the Eddington limit.

This effect would be further amplified in M$_{\star}<$10$^{10}$~M$_{\odot}$ galaxies, albeit it could be partly mitigated by assuming quite massive BH seeds (M$_{\rm BH, SEED}\gtrsim$ 10$^{5}$~M$_{\odot}$), and accounting for the minor contribution of low-M$_{\star}$ galaxies to the integrated SMBH accretion density.

In order to alleviate the deviation from the canonical M$_{\rm BH}$/M$_{\star}$ = 1/500, a lower radiative efficiency $\epsilon<$0.1 could be postulated, though insufficient to reconcile the two trends (see \citealt{Delvecchio+2019} for a detailed discussion). Alternatively, avoiding super-Eddington BH accretion would require the sBHAR distributions to peak at much lower L$_{\rm X}$/M$_{\star}$ with decreasing M$_{\star}$, inconsistent with current observables (e.g. \citealt{Aird+2019}). Shedding light on these issues would be relevant for a number of studies focusing on the evolution of $\lambda_{\rm EDD}$ distributions (e.g. \citealt{Aird+2012}; \citealt{Bongiorno+2012}; \citealt{Weigel+2017}; \citealt{Caplar+2018}; \citealt{Bernhard+2018}, \citeyear{Bernhard+2019}; \citealt{Aird+2019}; \citealt{Grimmett+2019}). 

We thus argue that testing this empirical prediction in the intrinsic Eddington-ratio space is essential to constrain the buildup of the M$_{\rm BH}$/M$_{\star}$ relation and standard prescriptions for BH accretion distributions. We propose that a highly complete sample of AGN above a certain Eddington ratio (obtained via reliable M$_{\rm BH}$ measurements) would be useful to observationally tie down the average break $\lambda_{\rm EDD, INTR}$ in a statistical manner.

\section{Summary and conclusions} \label{summary}

In this paper we decipher the evolution of the AGN duty cycle in galaxies from the XLF, separating the contribution of MS and SB galaxies since z$\sim$3. While these populations are known to display profound differences in structure and gas content, still open is the question of whether the rate and incidence of SMBH accretion depend on MS offset, and how they evolve over cosmic time. In order to account for the stochasticity of AGN activity and mitigate possible selection effects, we modeled the XLF as the convolution between the galaxy M$_{\star}$ function and a large set of simulated $\lambda_{\rm EDD}$ distributions, as done in a number of previous works (e.g. \citealt{Bongiorno+2012}; \citealt{Aird+2012}; \citealt{Caplar+2015}; \citealt{Jones+2017}; \citealt{Weigel+2017}; \citealt{Bernhard+2018}). Conversely to most studies, we assumed a very simple modeling, characterised by M$_{\star}$--\textit{independent} $\lambda_{\rm EDD}$ parameters (slopes and break), normalised with M$_{\star}$--\textit{dependent} BHAR/SFR relations reported in the literature (Section~\ref{method}). This allows us to derive a large set of predicted XLF, separately between MS and SB galaxies, with a simple statistical approach and well anchored to empirical grounds.

Our analysis relies on three prior assumptions (Section~\ref{assumptions}): (i) the XLF is predominantly made by MS and SB galaxies, while passive systems have a negligible contribution (as later confirmed in Section \ref{qxlf}); (ii) we parametrise the $\lambda_{\rm EDD}$ distribution as a broken powerlaw with slopes ($\alpha$,$\beta$) that meet at the break $\lambda^{*}$; (iii) the values of $\alpha$ and $\beta$ are assumed not to differ between MS and SB galaxies, at fixed redshift. 

The comparison between our model predictions and the observed XLF \citep{Aird+2015} reveals a very good agreement at all redshifts (Section~\ref{xlf}), which leads us to the following main results.

(1)~~We reproduce the observed XLF through a continuous flattening of the faint-end $\lambda_{\rm EDD}$ distribution, as well as a positive shift of the break $\lambda^{*}$ with redshift, consistently with previous studies (\citealt{Caplar+2015}, \citeyear{Caplar+2018}). Driven by our empirically-motivated assumptions, SB galaxies stand above by a constant offset of 0.8~dex, reaching break $\lambda^{*}$ close or slightly above the Eddington limit (Section~\ref{par_evo} and Figure~\ref{fig:par_evo}). 

(2)~~By splitting the XLF into M$_{\star}$ bins, we find that the bulk XLF is made by massive galaxies (10$^{10}<$M$_{\star}<$10$^{11}$~M$_{\odot}$) on the MS, while merging-driven BH accretion in SB galaxies becomes dominant only in bright quasars with L$_{\rm X}>$10$^{44.36}\cdot$(1+z)$^{1.28}$~erg~s$^{-1}$ (Fig.~\ref{fig:par_evo2}). The inferred BHARD traced by the MS population shows a peak at z$\sim$2, and declines at lower redshifts in a similar fashion with the SFRD \citep{Madau+2014}. Quiescent galaxies are estimated to contribute $<$6\% of the integrated BHARD at each redshift (Section \ref{bhard}).

(3)~~We underline that a M$_{\star}$-dependent relation between BHAR and SFR is strongly favored by our modeling, and in line with recent studies \citep{Aird+2019}. The best solution corresponds to BHAR/SFR $\propto$ M$_{\star}^{0.73[+0.22,-0.29]}$, while a constant BHAR/SFR trend is rejected at $\sim$3$\sigma$ significance, because would overpredict the XLF at low L$_{\rm X}$ arising from low-M$_{\star}$ galaxies (see \citealt{Bernhard+2018}). This finding implies that the cosmic buildup of SMBH and galaxy mass does not occur in lockstep at all epochs, but it evolves non linearly as the galaxy grows in M$_{\star}$ (Section~\ref{fgas}). 

(4)~~Our modeling successfully reproduces the relatively flat L$_{\rm X}$--SFR relation observed in X-ray selected AGN \citep{Stanley+2015} since z$\sim$3. This bolsters the reliability of our approach in predicting realistic SFR estimates for X-ray AGN across a wide L$_{\rm X}$ and redshift range (Section~\ref{testing}).

(5)~~Finally, we argue that the probability of finding highly-accreting ($\lambda_{\rm EDD}>$ 10\%) AGN notably increases with redshift, from 0.4\% (3.0\%) at z=0.5 to 6.5\% (15.3\%) at z=3 for MS (SB) galaxies (Fig.~\ref{fig:par_evo2}), which supports a longer AGN duty cycle in the early Universe, especially in dusty starbursting galaxies (Section~\ref{dutycycle}). This is expectable if the level of SMBH accretion is tightly linked to the amount of usable cold gas in the host. 

Concluding, our proposed framework serves as an important toy model for predicting the incidence of AGN activity in star-forming galaxies on and above the MS, the typical SFR and M$_{\star}$ of X-ray AGN, and the fraction of AGN lying within MS and SB galaxies, at different luminosities and cosmic epochs. This modeling also opens potential questions about super-Eddington BH growth and different $\lambda_{\rm EDD}$ prescriptions for explaining the assembly of the M$_{\rm BH}$--M$_{\star}$ relation.

Our key results broadly support a long-lasting interplay between SMBH accretion and star formation in galaxies, both showing enhanced activity at earlier epochs. This scenario is plausible if the evolution of cold gas content drives the triggering and maintenance of both phenomena over cosmic time. We speculate that merger- (or massive cold gas inflow-) driven SMBH accretion might be widespread in high-redshift (z$>$1) SB galaxies, explaining the onset of Eddington-limited activity and the longer AGN duty cycle relative to MS analogs.

\acknowledgments{ID is grateful to Iary Davidzon for useful input on the galaxy stellar mass function. ID is supported by the European Union's Horizon 2020 research and innovation program under the Marie Sk\l{}odowska-Curie grant agreement No 788679. AC acknowledges the support from grant PRIN MIUR 2017. RC acknowledges financial support from CONICYT Doctorado Nacional N$^\circ$\,21161487 and CONICYT PIA ACT172033. FV acknowledges financial support from CONICYT and CASSACA through the Fourth call for tenders of the CAS-CONICYT Fund, and CONICYT grants Basal- CATA AFB-170002.}

\bibliographystyle{apj}



\end{document}